\documentclass[aps,prd,reprint,preprintnumbers,superscriptaddress,nofootinbib,longbibliography,floatfix]{revtex4-1}

\usepackage[utf8]{inputenc}
\usepackage[T1]{fontenc}
\usepackage{lmodern}
\usepackage{microtype}
\usepackage{mathtools}

\usepackage{placeins}
\usepackage[caption=false]{subfig}
\usepackage{amsfonts}
\usepackage{graphicx}
\usepackage{hyperref}
\hypersetup{
  colorlinks=true,
  citecolor=blue,
  linkcolor=blue,
  urlcolor=blue
}

\usepackage{verbatim}
\usepackage{amsmath}
\usepackage{physics}
\usepackage{amssymb}
\usepackage{multirow}
\usepackage{slashed}
 
\usepackage{color}
\definecolor{mit-red}{rgb}{0.64,.12,0.2}
\definecolor{darkred}{rgb}{1.0,0.1,0.1}
\definecolor{darkgreen}{rgb}{0.1,0.7,0.1}
\definecolor{darkblue}{rgb}{0.1,0.1,1.0}

\usepackage{amsmath}
\DeclareMathOperator*{\argmax}{argmax}

\DeclareRobustCommand{\Sec}[1]{Sec.~\ref{sec:#1}}

\DeclareRobustCommand{\Tab}[1]{Table~\ref{tab:#1}}

\DeclareRobustCommand{\Fig}[1]{Fig.~\ref{fig:#1}}
\DeclareRobustCommand{\Figs}[2]{Figs.~\ref{fig:#1} and \ref{fig:#2}}
\DeclareRobustCommand{\Eq}[1]{Eq.~(\ref{eq:#1})}
\DeclareRobustCommand{\Eqs}[2]{Eqs.~(\ref{eq:#1}) and (\ref{eq:#2})}
\DeclareRobustCommand{\Reference}[1]{Ref.~\cite{#1}}

\def\reco{\text{reco}}

\begin{document}

\title{Seeing Double: Calibrating Two Jets at Once
}

\preprint{MIT-CTP  5680}

\author{Rikab Gambhir}
\email{rikab@mit.edu}
\affiliation{Center for Theoretical Physics, Massachusetts Institute of Technology, Cambridge, MA 02139, USA}
\affiliation{The NSF AI Institute for Artificial Intelligence and Fundamental Interactions}

\author{Benjamin Nachman}
\email{bpnachman@lbl.gov}
\affiliation{Physics Division, Lawrence Berkeley National Laboratory, Berkeley, CA 94720, USA}
\affiliation{Berkeley Institute for Data Science, University of California, Berkeley, CA 94720, USA}

\begin{abstract}

Jet energy calibration is an important aspect of many measurements and searches at the LHC. 
Currently, these calibrations are performed on a per-jet basis, i.e. agnostic to the properties of other jets in the same event. 
In this work, we propose taking advantage of the correlations induced by momentum conservation between jets in order to improve their jet energy calibration.
By fitting the $p_T$ asymmetry of dijet events in simulation, while remaining agnostic to the $p_T$ spectra themselves, we are able to obtain correlation-improved maximum likelihood estimates.
This approach is demonstrated with simulated jets from the CMS Detector, yielding a $3$-$5\%$ relative improvement in the jet energy resolution, corresponding to a quadrature improvement of approximately 35\%. %in 2011 CMS Open Simulation data. 
%
%This is possible without making any assumptions on the individual distributions of the jet momenta, only their momentum asymmetry, and does not introduce any additional bias when the two jets are studied inclusively.

\end{abstract}

%\date{\today}
\maketitle

{
\tableofcontents
}

\section{Introduction}

Jet energy calibration~\cite{collaboration_2011,CMS:2016lmd, ATLAS:2019oxp, ATLAS:2020cli, CMS-DP-2020-019, ATLAS:2023tyv} is a crucial ingredient of most measurements and searches at the Large Hadron Collider (LHC). 
Given a spray of dozens to hundreds of detected particles, the goal is to estimate the underlying pre-detector total energy or $p_T$ of the jet, along with a resolution on this estimate.
This is a complex and high-dimensional problem, and there have been a large number of proposals, many using machine learning (ML), to improve jet energy calibration~\cite{Komiske:2017ubm,ArjonaMartinez:2018eah,ATL-PHYS-PUB-2018-013,Martinez:2018fwc,Haake:2018hqn,Haake:2019pqd,CMS:2019uxx,ATL-PHYS-PUB-2019-028,ATL-PHYS-PUB-2019-028,Carrazza:2019efs,ATL-PHYS-PUB-2020-001,Baldi:2020hjm,Kasieczka:2020vlh, Maier:2021ymx, Gambhir:2022gua, Gambhir:2022dut, Li:2022omf, CRESST:2022qor, Gouskos:2022gjg, Kim:2023koz, Lieret:2023aqg, Holmberg:2023rfr} and uncertainty estimates~\cite{Sirunyan:2019wwa,Cheong:2019upg,Bollweg:2019skg,Bellagente:2021yyh,Araz:2021wqm,Kronheim:2021hdb}. 
Many of these approaches involve better or more sophisticated modeling of detector interactions or extracting more information from the high-dimensional particle phase space.

In this paper, we present an alternate method for improving jet energy calibrations and resolutions by considering information \emph{between} jets, rather than correlations \emph{within} jets.
In particular, we can obtain ``correlation-improved'' maximum likelihood estimates (MLE) and resolutions for jets in dijet events by using the fact their momenta add to approximately zero in the center-of-mass frame.
This can be done in a $p_T$ spectrum prior-independent way -- that is, without any assumptions on the overall $p_T$ distribution of the individual jets.
%calibration
We only assume that the momentum asymmetry between the two is either a Gaussian or symmetrized exponential of width $\Lambda$ that can be extracted from simulation. 
For jets with original resolution $\sigma$, this leads to calibrations with resolutions improved by an average of $\frac{\sigma^2}{2\Lambda^2} \sim 3$-$5\%$ (representing a 35\% improvement in quadrature) as estimated using simulations of the early Run 1 CMS Detector~\cite{cernopendata}, all while remaining unbiased.

While multi-jet information is not currently used to constrain the detector response of individual jets, constraints from multi-jet systems do play a role in jet calibrations. 
In particular, momentum conservation is used to determine how well the jet calibration is modeled by simulation, resulting in residual calibrations~\cite{collaboration_2011,CMS:2016lmd, ATLAS:2019oxp, ATLAS:2020cli, CMS-DP-2020-019, ATLAS:2023tyv,ATLAS:2018fwq,CMS:2023ebf}.
This information is the same as we use in this paper, only instead of constraining all jets at once, we constrain each jet individually.
For example, there are existing studies to use dijet balances to fit a $p_T$- and rapidity-dependent scale factor for the jet energy, but a given jet is not adjusted based on the energy of the balancing jet.
%Only the overall spectrum of the energy balance is used to derive the correction.  While we focus on dijets in this paper, the methodology would equally apply to multijet events, events with a jet balancing a non-jet, and events without any jets (e.g. lepton pairs).  The more precise the energy (or other kinematic) constraint, the larger the potential gains from our approach.  

In \Sec{statistics}, we go over the statistical methodology of the correlation-improved MLE.
In \Sec{example_toy}, we demonstrate our method on a simplified toy model to show its expected behavior. 
In \Sec{example_qcd}, we show how correlation-improved MLE may be used to improve the $p_T$ calibration of dijet events using simulated jets from the CMS Detector.
Finally, in \Sec{example_top}, we briefly discuss how this technique may be applied to dijets sourced from boosted resonances.
We present our conclusions and outlook in \Sec{conclusion}.

\section{Correlation-Improved Maximum Likelihood Estimates}\label{sec:statistics}

In the following, $x$ will represent what we can measure and $z$ will represent the true (`latent') values that we are trying to predict.\footnote{To simplify notation, we will use lowercase letters $x$ and $z$ to refer to both the random variable and the realization of the random variable.}
Given a single measurement $x \in \mathbb{R}^M$, we can construct a maximum likelihood estimate $\hat{z}$ of a latent parameter $z \in \mathbb{R}^N$ and its resolution in the Gaussian approximation $\sigma_{\hat{z}}$ as follows (see Ref.\ \cite{Gambhir:2022dut} for a review):
\begin{align}
    \hat{z}(x) &= \argmax_z L(x|z), \label{eq:MLE_mean}\\
    % \left[\sigma^2_{\hat{z}}(x)\right] &= -\frac{1}{2}\left[\frac{\partial^2 L(x|z)}{\partial z_i \partial z_j}\right]^{-1}\biggr\rvert_{z = \hat{z}(x)}, \label{eq:MLE_std}
    \left[\sigma^2_{\hat{z}}(x)\right]_{ab} &= -\left(\frac{\partial^2 L(x|z)}{\partial z_a \partial z_b}\right)^{-1}\biggr\rvert_{z = \hat{z}(x)}, \label{eq:MLE_std}
\end{align}
where $L(x|z)$ is shorthand for $\log(p(x|z))$.
The log-likelihood $L$ can be estimated from data or simulation, using ML techniques if either $x$ or $z$ is high-dimensional. 
The prototypical example is to think of $L(x|z)$ as a detector model or noise model, where noise or information loss is applied to a ``true'' parameter $z$ to produce the measured value $x$.
These estimates are manifestly independent of the true latent parameter prior $p(z)$, and produce independent estimates $\hat{z}$ for independent measurements $x$, which is a desirable feature for calibration.

Now, suppose that we measure a set of measurements all at once, with $x_1, ..., x_n$ (denoted $\Vec{x}$) corresponding to $z_1, ..., z_n$ (denoted $\Vec{z}$).  When each of the $n$ objects is isolated\footnote{The general calibration problem would allow for the dimensionality of $\Vec{x}$ to not necessarily equal that of $\vec{z}$.  In that case, the calibration is not universal and other approaches are required.}, $L(x_i|z_i)$ is independent of $i$, i.e. the detector adds noise to each latent parameter independently.
If the $z_i$ are uncorrelated, then the total likelihood for the entire measurement factorizes, and we can obtain independent maximum likelihood estimates and resolutions for each $z_i$.
However, suppose we had prior knowledge of correlations $p(z_1, ... z_n)$. 
Then, by Bayes' rule, 
\begin{align}
    L(\Vec{z}|\Vec{x}) &= L(\Vec{x}|\Vec{z}) + L(\Vec{z}) - L(\Vec{x}), \nonumber\\
    &= \sum_{i} L(x_i | z_i) + L(\Vec{z}) + \text{Norm}\label{eq:bayes}.
\end{align}

The left hand side of \Eq{bayes} is dependent on the prior, $L(\Vec{z})$, which is undesirable for calibration tasks. 
However, it is sufficient for the multi-measurement calibrations we aim to perform that we are only independent with respect to each individual $L(z_i)$ -- that is, while we want to maintain our prior knowledge of the correlations between each $z_i$ (e.g. ``How is object 1 related to object 2?''), we do not want our inference to depend on the distribution of $z_i$ (e.g. ``How many of object 1 are there?'').

One way this can be accomplished is by decomposing $L(\Vec{z}) = L(z_n|z_{n-1} ... z_1) + L(z_{n-1}|z_{n-2}...z_1) + ... + L(z_2|z_1) + L(z_1)$. 
The entire prior dependence  is placed on $L(z_1)$, and so we can isolate and remove it to construct a likelihood.

Having removed explicit per-parameter prior dependence, we can construct a likelihood $L'(\vec{x}|\vec{z})$ (ignoring unimportant normalizations):
\begin{align}
    L'(\Vec{x}|\Vec{z}) &= \sum_i \left(L(x_i | z_i) + L(z_i|z_{i-1}...z_1)\right), \label{eq:general_likelihood}
\end{align}
which can be thought of as the ordinary likelihood $L(\vec{x}|\vec{z})$ plus a constraint on the relationship between the $z_i$'s.
The prime on $L'(\vec{x}|\vec{z})$ is to indicate that this likelihood is \emph{not} just $L(\vec{x}|\vec{z})$, as it contains addition information about the constraints.
We emphasize that this is not completely prior independent: The likelihood $L'$ does depend on $p(\vec{z})$, but only through the correlations between the $z_i$'s and not directly from any individual $p(z_i)$.

From \Eq{general_likelihood}, it is now possible to achieve a correlation-improved MLE estimate and resolution using \Eqs{MLE_mean}{MLE_std}.
Typically, as in the case in jet physics applications, $x$ is very high dimensional (e.g. $M\sim \mathcal{O}(10^3)$), whereas $z$ is often relatively low dimensional (e.g. for only the jet energy, $N=1$).
We can attempt to simplify the calculation of \Eq{general_likelihood} by recasting it in terms of the low-dimensional $\hat{z}(x)$, as calculated in \Eq{MLE_mean}: if $\hat{z}(x)$ is a (nearly) sufficient estimator of $z$\footnote{This is not guaranteed -- prior independence is necessary, but not sufficient to be unbiased, as discussed in Ref.\ \cite{Gambhir:2022dut}. However, if the noise model $L(x|z)$ is Gaussian, $\hat{z}$ will indeed be unbiased.}, then we may write:
\begin{align}
    L(\hat{z}(x) | z) &\approx L(x | z), \\
    &\approx \log(\mathcal{N}(z, \sigma^2_{z})) ,
\end{align}
where $\mathcal{N}$ is the probability density corresponding to the normal distribution with mean $z$ and variance $\sigma_z^2$, and $\sigma_z$ is the resolution of the estimator $z(x)$ (in most cases, $\sigma_z$ is the added Gaussian noise).
Therefore, we can write:
\begin{align}
    L'(\Vec{x}|\Vec{z}) &= \sum_i \left(L(\hat{z}_i(x_i) | z_i) + L(z_i|z_{i-1}...z_1)\right), \label{eq:general_likelihood_z}
\end{align}
which allows us to us our previously-determined per-object estimators as inputs to the correlation-improved estimator.

\subsection{The $n = 2$ Gaussian  Case}\label{sec:gaussian}

We are primarily interested in the $n = 2$ case of \Eq{general_likelihood_z}, where we have two correlated objects $z_1$ and $z_2$. 
The correlation is completely determined by $L(z_2|z_1)$, which we will assume takes the form:
\begin{align}
    L(z_1|z_2) &= L(z_1 - z_2).
\end{align}
We will assume that $z_1 - z_2$ follows a Gaussian distribution with mean 0, though much of what we say here qualitatively holds for most other non-pathological distributions. 
In other words, $z_1$ equals $z_2$, up to Gaussian noise $\Lambda$:
\begin{align}
    L(z_1 - z_2) = -\frac{1}{2\Lambda^2}(z_1-z_2)^2 + \text{Norm}.
\end{align}
Then, given a set of measurements $x_1$ and $x_2$, we can find correlation improved estimates of the latent parameters $z_1$ and $z_2$ (which we denote $\hat{z}_1'(x)$ and $\hat{z}_2'(x)$) in terms of the original MLE estimates and resolutions (unprimed) by applying \Eq{MLE_mean} to \Eq{general_likelihood_z}: 
\begin{align}
    \hat{z}'_1(x_1, x_2) &= \frac{1}{\sigma_\Lambda^2}\left(\hat{z}_1 (\Lambda^2 + \sigma_{\hat{z}_2}^2) + \hat{z}_2\sigma_{\hat{z}_1}^2 \right), \nonumber\\
    \hat{z}'_2(x_1, x_2) &= \frac{1}{\sigma_\Lambda^2}\left(\hat{z}_2(\Lambda^2 + \sigma_{\hat{z}_1}^2) + \hat{z}_1\sigma_{\hat{z}_2}^2 \right), \label{eq:mean2}
\end{align}
where for convenience we have defined $\sigma_\Lambda^2 \equiv \Lambda^2 + \sigma_{\hat{z}_1}^2 + \sigma_{\hat{z}_2}^2$.
We have also suppressed the dependence of $\hat{z}$ and $\sigma_{\hat{z}}$ on $x_1$ and $x_2$ for notational cleanliness.

These improved estimates are completely unbiased, \emph{as long as $x_1$ and $x_2$ are unordered}.
More precisely, as long as we are agnostic to whether or not $x_1 > x_2$, then the bias $b(z)$:
\begin{align}
    b(z_{1,2}) &= \mathbb{E}_{p_{\rm Test}}[\hat{z}'_{1,2}(x_1, x_2) - z_{1,2}] \nonumber,\\
    &= 0 \label{eq:bias_0},
\end{align}
where $p_{\text{Test}}$ is any test set such that $p(z_2 | z_1)$ is correlated with the same Gaussian $\Lambda$.\footnote{This is not strictly necessary -- many other choices of the distribution of $|z_1 - z_2|$, including Gaussians and symmetrized exponentials for \emph{any} width $\Lambda$, also lead to zero bias }
This is no longer true if we know which $x$ is ``leading'' and ``subleading'' -- the estimates $\hat{z}'$ will tend towards the resolution-weighted average of $x_1$ and $x_2$, so the leading estimate always decreases and the subleading estimate always increases.

We can also compute the corresponding resolutions of each estimator using \Eq{MLE_std}, $\sigma_{\hat{z}_1}'(x)$ and $\sigma_{\hat{z}_2}'(x)$, as the diagonal elements of the following covariance matrix:
\begin{align}        
    \Sigma'^2_{\hat{z}}(x_1, x_2) &= \frac{1}{\sigma_\Lambda^2}\begin{bmatrix}
        \sigma_{\hat{z}_1}^2(\sigma_\Lambda^2 - \sigma_{\hat{z}_1}^2) & \sigma_{\hat{z}_1}^2 \sigma_{\hat{z}_2}^2 \\
        \sigma_{\hat{z}_1}^2 \sigma_{\hat{z}_2}^2 &\sigma_{\hat{z}_2}^2(\sigma_\Lambda^2 - \sigma_{\hat{z}_2}^2)
    \end{bmatrix}\label{eq:var2}
    \nonumber\\
    &\downarrow\nonumber\\
    \sigma'^2_{\hat{z}_1}(x_1, x_2) &= \sigma_{\hat{z}_1}^2\left(1-\frac{\sigma_{\hat{z}_1}^2}{\sigma_{\hat{z}_1}^2 + \sigma_{\hat{z}_2}^2 + \Lambda^2}\right)\nonumber\\
    \sigma'^2_{\hat{z}_1}(x_1, x_2) &= \sigma_{\hat{z}_2}^2\left(1-\frac{\sigma_{\hat{z}_2}^2}{\sigma_{\hat{z}_1}^2 + \sigma_{\hat{z}_2}^2 + \Lambda^2}\right).
\end{align}
In the case that $z_1$ or $z_2$ are multi-dimensional, this covariance matrix should be thought of as block-diagonal in $z_1$-$z_2$ space.
%
% \begin{align}        
%     \frac{1}{\sigma'^{2}_{\hat{z}_1}(x_1, x_2)} &= \frac{1}{\sigma^{2}_{\hat{z}_1}(x_1)} + \frac{1}{\Lambda^2}, \nonumber\\
%     \frac{1}{\sigma'^{2}_{\hat{z}_2}(x_1, x_2)} &= \frac{1}{\sigma^{2}_{\hat{z}_2}(x_2)} + \frac{1}{\Lambda^2}, \label{eq:var2}
% \end{align}

% Next, we consider the case where $z_1 - z_2$ is exponentially distributed:
% %
% \begin{align}
%     L(z_1 - z_2) = -\frac{1}{\Lambda}|z_1 - z_2| + \text{Norm}. 
% \end{align}
% %
% The analysis becomes more complicated as the likelihood is not differentiable at 0\footnote{We could have considered a 1-sided exponential where $z_1 > z_2$, but this does not solve the issue, as the relative ordering of $z_1$ and $z_2$ may shift under noise, again encoded by the singularity at 0.}. We can extract three possible cases for the maximum likelihood values of $z_1$ and $z_2$, corresponding to whether $z_1 > z_2, z_1 < z_2$, or $z_1 = z_2$:
% %
% \begin{align}
%     \textbf{Case 1 \& 2}: \hat{z}'_1(x_1, x_2) &= \hat{z}_1(x_1) \mp \frac{\sigma^2_{\hat{z}_1}(x_1)}{\Lambda},\nonumber\\
%     \hat{z}'_2(x_1, x_2) &= \hat{z}_2(x_2) \pm \frac{\sigma^2_{\hat{z}_2}(x_2)}{\Lambda},\nonumber\\
%     \textbf{Case 3}: \hat{z}'_1(x_1, x_2) &= \hat{z}'_2(x_1, x_2) = \frac{\sigma_{\hat{z}_1}^2 \hat{z}_2 + \sigma_{\hat{z}_2}^2 \hat{z}_1}{\sigma_{\hat{z}_2}^1+\sigma_{\hat{z}_2}^2}.
% \end{align}
% %
% In order to compute the true MLE estimate, one would have to compute the explicit likelihood given each of these three cases and select the maximum.

%
%
%
For any $\Lambda \geq 0$, \Eq{var2} always improves the resolutions of both measurements.
In the limit $\Lambda \to \infty$, we expect the correlation structure of $L(z_1 - z_2)$ to vanish, and indeed we see there is no improvement to the resolution and the off-diagonals of the covariance matrix disappear.
In this case, the improved estimates $\hat{z}'$ are completely unchanged from the original estimates $\hat{z}$.
On the other hand, in the limit that $\Lambda \to 0$, we expect that $z_1 = z_2$ exactly, and indeed $\hat{z}_1' = \hat{z}_2'$ in this limit.
Both estimates converge to the resolution-weighted average of $\hat{z}_1$ and $\hat{z}_2$, with the improved resolution approaching the usual inverse-quadrature sum of $\sigma_{\hat{z}_1}$ and $\sigma_{\hat{z}_2}$.
In this limit, the two estimates become perfectly correlated with a Pearson correlation coefficient of $\rho = 1$.
In the case where $\sigma_{\hat{z}_1} \sim \sigma_{\hat{z}_2}$, this leads to improved resolutions of $\sigma'_{\hat{z}} \sim \sigma_{\hat{z}_1} / \sqrt{2}$, an improvement of nearly $30\%$.

If we relax the assumption that $L(z_1 - z_2)$ is a Gaussian and instead consider generic unimodal and smooth distributions, it is still possible to obtain calibration-improved estimates. So long as $L(z_1 - z_2)$ has a sufficiently simple parametric form (which can be obtained from a simple fit) and is convex, it is still straightforward to calculate the correlation-improved maximum likelihood estimates and improved Gaussian resolutions, albeit only numerically.

\subsection{The $n = 2$ Exponential Case}\label{sec:exponential}

\begin{figure}[tbh]
    \centering
    \includegraphics[width=0.95\linewidth]{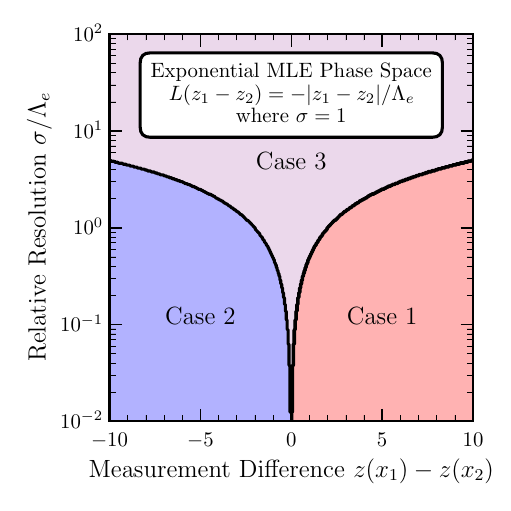}
    \caption{Phase space plot showing which of the three solutions presented in \Eq{exponential_mean} is preferred as a function of $x_1-x_2$ and $\sigma/\Lambda_e$, where $\sigma_{\hat{z}_1} = \sigma_{\hat{z}_2} = \sigma$ in this plot.}
    \label{fig:phase_space}
\end{figure}

The case where $L(z_1 - z_2)$ is a symmetrized exponential is pathological and qualitatively different than the Gaussian case.
If we take $z_1$ to be equal to $z_2$ up to \emph{exponential} noise $\Lambda_e$ of the form:
\begin{align}
    L(z_1 - z_2) = -\frac{1}{\Lambda_e}|z_1 - z_2| + \text{Norm},
\end{align}
then the singular behavior of the distribution at $z_1 = z_2$ causes our maximum-likelihood estimates to be nonanalytic.
Indeed, we find that there are three possible critical values, corresponding to being on the left of, directly on, or to the right of the critical point $z_1 = z_2$:
\begin{align}
    \textbf{Case 1 \& 2}: \hat{z}'_1(x_1, x_2) &= \hat{z}_1(x_1) \mp^1_2 \frac{\sigma^2_{\hat{z}_1}(x_1)}{\Lambda_e},\nonumber\\
    \hat{z}'_2(x_1, x_2) &= \hat{z}_2(x_2) \pm^1_2 \frac{\sigma^2_{\hat{z}_2}(x_2)}{\Lambda_e},\nonumber\\
    \textbf{Case 3}: \hat{z}'_1(x_1, x_2) &= \hat{z}'_2(x_1, x_2) = \frac{\sigma_{\hat{z}_1}^2 \hat{z}_2 + \sigma_{\hat{z}_2}^2 \hat{z}_1}{\sigma_{\hat{z}_2}^1+\sigma_{\hat{z}_2}^2}. \label{eq:exponential_mean}
\end{align}
The only way to disambiguate the three cases is to evaluate the full likelihood (\Eq{general_likelihood_z}) then determine which of the three gives the maximum likelihood, which in general depends on the values of $\hat{z}_1$, $\hat{z}_2$, $\sigma_{\hat{z}_1}$, $\sigma_{\hat{z}_2}$, and $\Lambda_e$.
However, we can glean some insight by considering the simpler case where $\sigma_{\hat{z}_1} = \sigma_{\hat{z}_2} = \sigma$, in which case the full likelihoods become, up to unimportant normalization factors:
\begin{align}
    L_{1,2} &= -\left|\frac{\hat{z}_1-\hat{z}_2}{\Lambda_e} \mp \frac{2\sigma^2}{\Lambda_e^2}\right| - \frac{\sigma^2}{\Lambda_e^2},\\
    L_3 &= -\frac{(\hat{z}_1 - \hat{z}_2)^2}{4\sigma^2}.
\end{align}
If $\frac{(\hat{z}_1-\hat{z}_2)^2}{4\sigma^2} < \frac{\sigma^2}{\Lambda^2}$, then Case 3 is preferred -- otherwise, Case 1 or 2 is preferred, depending on the sign of $x_1 - x_2$. 
This can be understood as the estimators ``snapping'' to the resolution-weighted average if the measured values are sufficiently close, as controlled by $\Lambda_e$. 
This behavior is reminiscent of the similar ``snapping'' or ``lasso'' behavior of L1 regulators~\cite{51791361-8fe2-38d5-959f-ae8d048b490d}.\footnote{In fact, it is exactly the same -- L1 regulators can be thought of as placing a symmetrized exponential prior on parameters.}  
Otherwise, the improved estimators are just a slight shift of the original estimators.
In \Fig{phase_space}, we show which of the three solutions is preferred as a function of $x_1-x_2$ and $\sigma/\Lambda_e$.

Cases 1 and 2 merely shift the estimators, and thus the resolution is unchanged.\footnote{There is a slight change due to the dependence of $\sigma_{\hat{z}}$ on $x$, making the shift not entirely constant, but we will assume $\sigma$ varies slowly with $x$ and ignore it.}
Like the Gaussian case, as long as the measurements are unordered, these shifts do not produce bias, since the sign of the shift will average to zero.
On the other hand, Case 3 produces the ordinary resolution-weighted average of the two measurements, and thus the improved resolution is the ordinary inverse-quadrature sum of the original resolutions.
That is:
\begin{align}
    \sigma'^2_{\hat{z}_i} = \begin{cases} 
       \sigma^2_{\hat{z}_i} & \text{Cases 1 \& 2} \\
      (\sigma^{-2}_{\hat{z}_1} + \sigma^{-2}_{\hat{z}_2})^{-1} & \text{Case 3} \label{eq:exponential_var2}
   \end{cases}.
\end{align}

Note that while it appears that Case 3 of \Eq{exponential_mean} and all of \Eq{exponential_var2} are explicitly independent of $\Lambda_e$, there is still a strong implicit dependence on $\Lambda_e$ through the selection of the three cases.
In the limits $\Lambda_e \to \infty$ and $\Lambda_e \to 0$, these estimates and resolutions agree exactly with the Gaussian version. 
This was to be expected, as the two limits represent the no-information and perfect-correlation cases respectively. 
However, unlike the Gaussian case, where \Eq{var2} does \emph{not} depend directly on $\hat{z}_1$ or $\hat{z}_2$, \Eq{exponential_var2} \emph{does} depend on $\hat{z}_1$ and $\hat{z}_2$ through the case choice.
In the case where $\sigma_{\hat{z}_1} = \sigma_{\hat{z}_2} = \sigma$ for simplicity, we can use the exponential prior on $|z_1 - z_2|$ to directly obtain the average improvement in resolution:
\begin{align}
    \sigma'^2_{avg} &= \sigma^2\left[\frac{1}{2} + \frac{1}{2}e^{-2\sigma^2/\Lambda_e^2}\right] \nonumber\\
    &\approx \sigma^2\left(1 - \frac{\sigma^2}{\Lambda_e^2}\right).\label{eq:large_lambda}
\end{align}
Note that this lines up with \Eq{var2} in the limit that $\Lambda \gg \sigma$.
That is, even though the Gaussian and exponential priors have very different behavior, we expect that they should both have the same \emph{average} correlation-improved resolution when $\Lambda = \Lambda_e \gg \sigma$.

\section{Example 1: Toy Model Decay}\label{sec:example_toy}

We begin with an example of applying the correlation-improved MLE using known distributions where we have full analytic and numeric control of the analysis. 

Suppose we have a simplistic toy model where events have ``final state'' consisting of three (distinguishable) particles, two of which are visible and one invisible. 
Focusing only on the $x$ component of momentum, we measure the momenta of the fist two particles to be $k_{1, \reco}$ and $k_{2, \reco}$, with unknown true values $k_{1}$ and $k_{2}$ that we would like to estimate.
Assume that our detector adds a known amount of Gaussian noise $\sigma_1$ and $\sigma_2$ to each of the two measured particles, and that we cannot measure the third particle. 
We will also assume that we know that the true momentum $k_3$ follows a Gaussian with mean 0 and variance $\Lambda^2$. 
This cartoon model resembles a $\tau \to \pi^-\pi^0 \nu_\tau$ decay in the rest frame of the $\tau$, where $\sigma_- \neq \sigma_0$, and will enable us to demonstrate the features of the correlation-improved MLE. 
We will revisit the assumption of knowing the parent particle is perfectly at rest in \Sec{example_top} in the case of boosted resonance decays.

We can independently apply maximum likelihood estimation to obtain estimates for the values and resolutions of each measurement:
\begin{align}
    \hat{k}_{1} &= k_{1, \reco},\quad\hat{\sigma}_{1} = \sigma_{1},\nonumber\\
    \hat{k}_{2} &= k_{2, \reco},\quad\hat{\sigma}_{2} = \sigma_{2}. \label{eq:toy_MLE}
\end{align}

Momentum conservation will induce correlations between $k_1$ and $k_2$, via $k_3$.
To match the sign conventions of \Sec{statistics}, we will choose $k_2$ to be in the negative direction.
Since $k_1 - k_2 + k_3 = 0$, these correlations take the form:
\begin{align}
    p(k_1 - k_2) = \mathcal{N}(0, \Lambda^2),\label{eq:gaussian_form}
\end{align}
which is precisely the setup required to apply \Eqs{mean2}{var2} to improve the estimates and resolutions obtained in \Eq{toy_MLE}, with $k_\reco$ corresponding to the measurement $x$ and $k$ corresponding to the parameter $z$.

\begin{figure*}[tp]
\centering
    \subfloat[]{
         \includegraphics[width=\columnwidth]{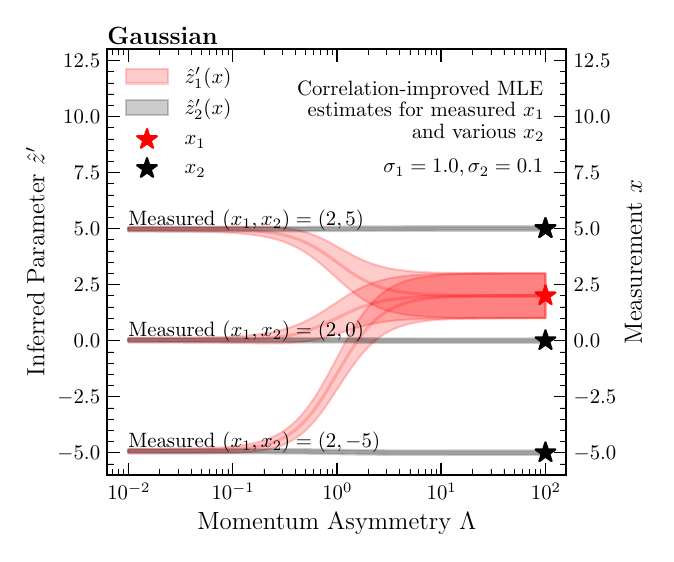}
        \label{fig:toy_example_MLE_0}
    }
    \subfloat[]{
        \includegraphics[width=\columnwidth]{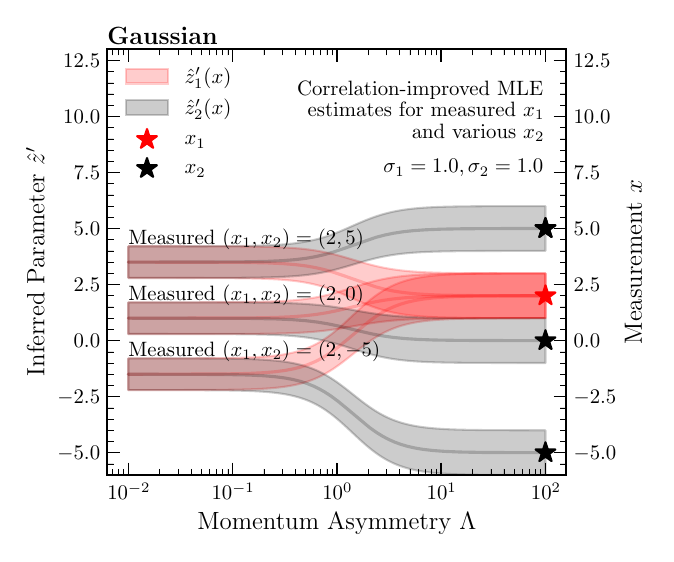}
        \label{fig:toy_example_MLE_1}
    }
    \caption{The Gaussian correlation-improved MLE parameters $\hat{z}'$ and $\sigma_{\hat{z}}'$ as a function of $\Lambda$, for $\sigma_1 = 1.0$ and $\sigma_2 = 0.1$ (a), and $\sigma_1 = 1.0$ and $\sigma_2 = 1.0$ (b). A single $x_1$ measurement is shown (red) and several possible corresponding $x_2$ measurements are shown (black). The stars correspond to the original measurement, and the bands correspond to the inferred parameter $z$ with its resolution.}
    \label{fig:toy_MLE}
\end{figure*}

In \Fig{toy_MLE}, we show the results of the correlation-improved MLE procedure (\Eqs{mean2}{var2}) for various choices of missing momentum scale $\Lambda$, resolutions $\sigma_{1,2}$, and measurements $x_{1,2}$. 
We pick two regimes: $\sigma_1 \gg \sigma_2$ (\Fig{toy_example_MLE_0}), and  $\sigma_1 \sim \sigma_2$ (\Fig{toy_example_MLE_1}) -- between these two plots, and the varying $\Lambda$ on the $x$ axis, all possible regimes in the $(\Lambda, \sigma_1, \sigma_2)$ parameter are represented.
We pick one specific value for the $x_1$ measurement (red) and several values of a possible $x_2$ measurement (black), where the value of the measurements are represented by stars. Then, for each $(x_1, x_2)$ pair, we draw bands corresponding to $\hat{z}'(x_1, x_2) \pm \sigma_{\hat{z}}'(x_1, x_2)$ as a function of $\Lambda$. We can see that as $\Lambda \to 0$, the estimates converge to the same resolution-weighted average of the two measurements. 
On the other hand, as $\Lambda \to \infty$, the improved resolutions are no different than those expected by \Eq{toy_MLE}.

\begin{figure*}[tbp]
\centering
    \subfloat[]{
         \includegraphics[width=\columnwidth]{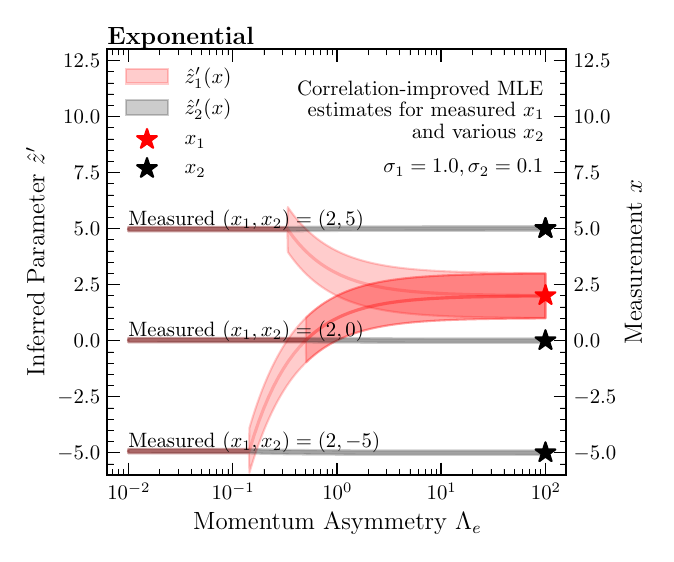}
        \label{fig:toy_example_MLE_0_EXPONENTIAL}
    }
    \subfloat[]{
        \includegraphics[width=\columnwidth]{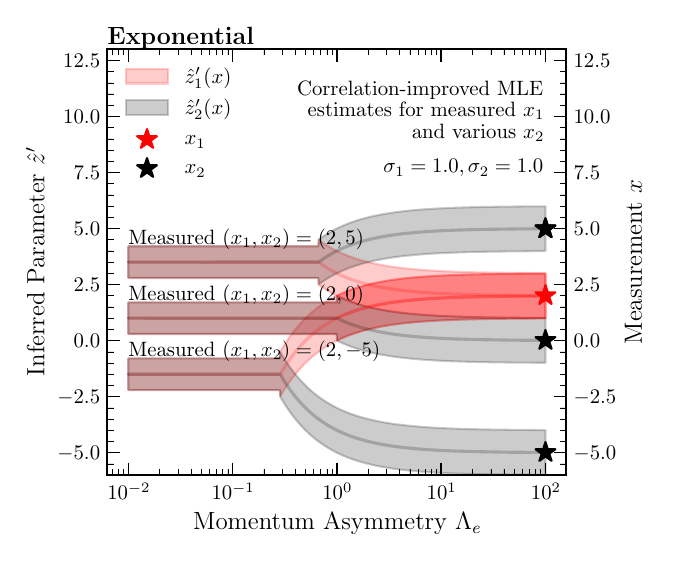}
        \label{fig:toy_example_MLE_1_EXPONENTIAL}
    }
    \caption{The same as \Fig{toy_MLE}, but with an exponential prior rather than a Gaussian prior. Note the non-smooth ``snapping'' behavior.}
    \label{fig:toy_MLE_EXPONENTIAL}
\end{figure*}

Similarly, we can also analyze the case of an exponential prior for $k_1 - k_2$: 
\begin{align}
    p(k_1 - k_2) = \frac{1}{2\Lambda_e}e^{-\frac{|k_1 - k_2|}{\Lambda_e}},\label{eq:exp_form}
\end{align}
using \Eqs{exponential_mean}{exponential_var2}.
In this scenario, the momentum of our cartoon $\nu$ particle is exponentially close to zero. 
The results of this calibration are shown in \Fig{toy_MLE_EXPONENTIAL}.
Note that while the limiting behavior of these curves are similar to the Gaussian \Fig{toy_MLE}, we can observe the characteristic ``snapping'' behavior of the exponential as $\Lambda \to \mathcal{O}(\sigma)$.

\section{Example 2: QCD Dijets}\label{sec:example_qcd}

We now apply our technique to a more realistic scenario: the calibration of jet energies at the LHC, using the CMS Detector~\cite{CMS:2008xjf} as an example. Our approach is deployed on simulated dijet events from CMS using early Run 1 conditions. 
In particular, in dijet events, we expect the momentum of the two jets to be correlated due to momentum conservation, which is precisely the scenario in which we can apply correlation-improved calibration.

\subsection{Dataset}

We use the same CMS Open Simulation~\cite{cernopendata, CMS:2008xjf} dataset as described in Ref.\ \cite{Gambhir:2022gua}, namely the CMS2011AJets simulation dataset.
This dataset consists of simulated dijets generated using \textsc{Pythia 6}~\cite{Sj_strand_2006} with a \textsc{Geant4}-based~\cite{AGOSTINELLI2003250} simulation of the CMS detector.
This dataset is formatted in the MIT Open Data (MOD) format -- more details about this format and this dataset can be found in Ref.\ \cite{komiske_patrick_2019_3340205}.

The data consists of both detector-level ``measured'' jets, SIM, and corresponding generator-level ``truth'' jets, GEN.
Each SIM event consists of a list of particle flow candidates (PFCs)~\cite{Beaudette:2013kbl}, which are the reconstructed four-momentum and particle identification (PID) for each measured particle.
The PFCs are clustered into anti-$k_t$ jets with a radius parameter of $R = 0.5$~\cite{Cacciari:2005hq,Cacciari_2008,Cacciari:2011ma}.
The dataset has an implicit SIM $p_T$ cut of 375 GeV on all jets, which was applied to avoid trigger turn-on affects. 
We additionally require that all jets are within $|\eta| < 2.4$ and that the jets are of at least ``medium'' quality~\cite{CMS:2010xta}.
Within this dataset, almost all (99.4\%) jets are paired up into dijet events.
The remaining non-paired jets are discarded. 
For each dijet event, we keep track of which is the leading and subleading jet at SIM level.
This information is \emph{only} used to bin the jets and to differentiate the jets in the forthcoming plots -- this is important for our calibrations to remain unbiased.

\begin{figure*}[tbh]
\centering
    \subfloat[]{
         \includegraphics[width=\columnwidth]{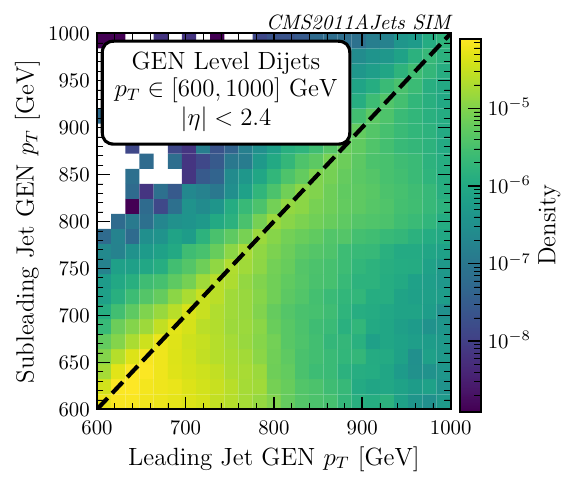}
        \label{fig:leading_subleading_GEN}
    }
    \subfloat[]{
        \includegraphics[width=\columnwidth]{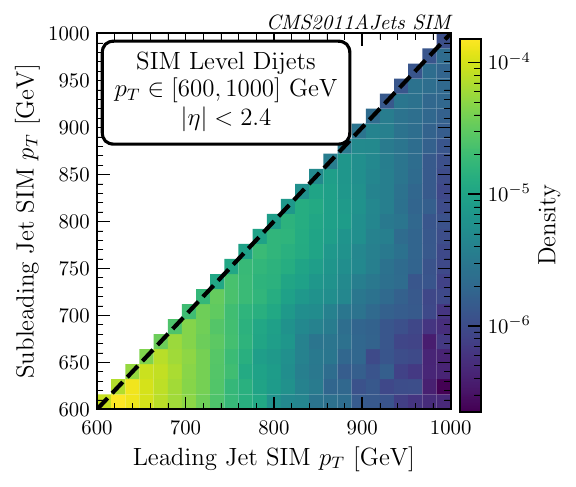}
        \label{fig:leading_subleading_SIM}
    }
    \caption{Distribution of the (a) GEN $p_T$ and the (b) SIM $p_T$ of the leading and subleading jets (as determined by the SIM $p_T$) of each event in the dataset.}
    \label{fig:leading_subleading}
\end{figure*}

To mitigate the effect of the global $375$ GeV trigger cut on all jets within this dataset, we only look at events whose leading jet $p_T$ is greater than 600 GeV.
In \Fig{leading_subleading}, we show the distribution of the leading and subleading $p_T$'s for each dijet event satisfying our cuts at both the SIM and GEN level. 
For the following studies, the latent variable of interest $z$ is the GEN jet $p_T$, while the corresponding measured quantities $x$ will be either the SIM jet $p_T$ or the SIM PFCs. 
We bin the dijet events according to their leading SIM $p_T$. 
The bins we consider are: $[600, 700]$ GeV, $[700, 800]$ GeV, $[800, 900]$ GeV, and $[900, 1000]$ GeV.
We perform our analysis separately on each bin, as well as one additional time on the entire inclusive $[600, 1000]$ GeV set.

\subsection{Momentum Asymmetry Fits}

Due to momentum conservation, the $p_T$'s of dijets are correlated.
In particular at truth level, $p_{T,1} \approx p_{T,2}$ -- however, a variety of factors make this inexact, such as such as initial/final state radiation, the underlying event, neutrinos/electroweak radiation, and inherent transverse momentum inbalances (see TMD effects~\cite{Boussarie:2023izj})

% In particular, at truth level , $p_{T,1} \approx p_{T,2}$ -- however, a variety of factors make this inexact, such as the presence of other softer jets~\cite{CMS:2013vbb, ATLAS:2020vup, Czakon:2021mjy}, extra radiation in the jet cone~\cite{Cacciari:2008gn, CMS:2020ebo}, missing transverse energy (MET)~\cite{Apresyan:2015kla, ATLAS:2018txj}, or TMD effects~\cite{Hautmann:2017fcj, BermudezMartinez:2021zlg,BermudezMartinez:2022bpj} \bn{I'd maybe would have said `...such as initial/final state radiation, the underlying event, neutrinos/electroweak radiation, and inherent transverse momentum inbalances (see TMD effects~\cite{})'.  I'm not sure we need cites for anything other than some review for the last one.}.
%
As in the case of the toy example presented in \Sec{example_toy}, as long as the distribution $L(p_{T,1} - p_{T,2})$ is known, it is possible to use  Eqs.\ \eqref{eq:mean2}-\eqref{eq:var2} to obtain improved calibrations beyond what is possible in the single-jet case.

We will perform two different fits and perform two independent analyses: one assuming an exponential fit with width $\Lambda_e$ with the same functional form as \Eq{exp_form}, and one assuming a Gaussian fit with width $\Lambda$ with the same functional form as \Eq{gaussian_form}.
We will find that the exponential fits more closely match the distribution, but since exponentials have pathological behavior (as discussed in \Sec{exponential}), we will do the analysis a second time with Gaussian fits.  Consistent improvements in the resolution are still possible even if the particle-level fit is not exact.  

%by closing our eyes and pretending that the distributions are Gaussian anyways for the sake of comparison.

\begin{figure*}[tbh]
\centering
    \subfloat[]{
         \includegraphics[width=\columnwidth]{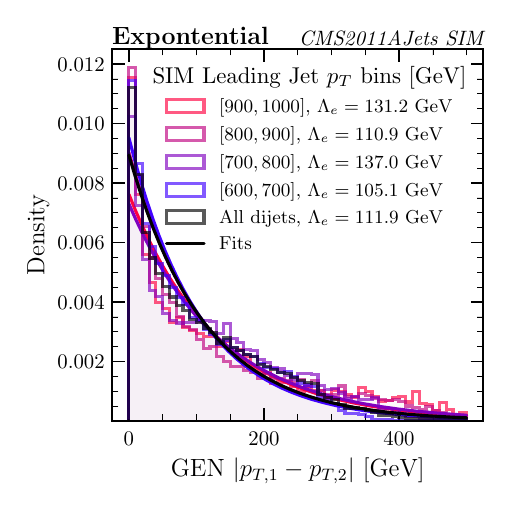}
    \label{fig:Lambda_exponential}
    }
    \subfloat[]{
        \includegraphics[width=\columnwidth]{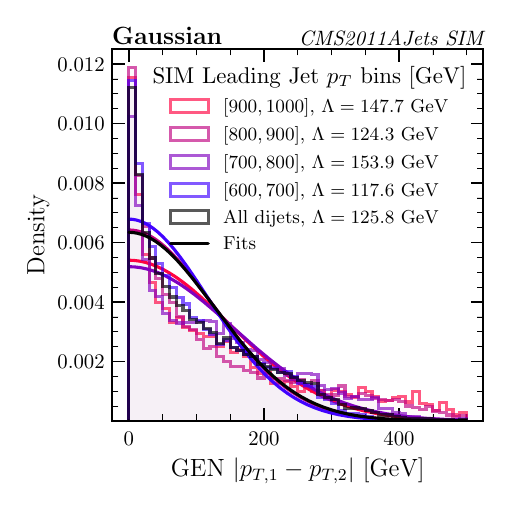}
        \label{fig:Lambda_gaussian}
    }
    \caption{Distribution of the difference in GEN $p_T$ between the leading and subleading jets in each event, for several leading $p_T$ bins. 
    The absolute value is taken for plotting convenience -- the labels $1$ and $2$ are random and the curves are symmetrized about the vertical axis.
    An (a) exponential and a (b)  Gaussian distribution is fit to each distribution, and the parameters $\Lambda_e$ and $\Lambda$ (respectively) are extracted to describe the extent to which momentum is conserved.}
    \label{fig:Lambda}
\end{figure*}

\begin{table*}[tpbh!]
\centering
\begin{tabular}{l|c|c|c|c|c|c}
Bin [GeV] & MSE (Exponential) & $\Lambda_e$ [GeV] & $\Lambda_e /$ Central $p_T$ & MSE (Gaussian) & $\Lambda$ [GeV] & $\Lambda /$ Central $p_T$ \\
\hline
$[600,700]$ & $2.8 \times 10^{-7}$ & 105.1 & 0.162 & $6.7 \times 10^{-7}$ & 117.6 & 0.181 \\
$[700,800]$ & $2.8 \times 10^{-7}$ & 137.0 & 0.183 & $8.7 \times 10^{-7}$ & 153.9 & 0.205 \\
$[800,900]$ & $2.9 \times 10^{-7}$ & 110.9 & 0.130 & $5.6 \times 10^{-7}$ & 124.3 & 0.146 \\
$[900,1000]$ & $2.2 \times 10^{-7}$ & 131.2 & 0.138 & $7.2 \times 10^{-7}$ & 147.7 & 0.155 \\
\hline
Inclusive & $2.7 \times 10^{-7}$ & 111.9 & 0.140 & $7.5 \times 10^{-7}$ & 125.8 & 0.157 \\
\hline
\end{tabular}
\caption{Mean squared errors and fitted $\Lambda$ values for the exponential and Gaussian fits to the $|p_{T,1} - p_{T,2}|$ distributions for different leading jet $p_T$ bins.}
\label{tab:fits}
\end{table*}

In \Fig{Lambda}, we show the distribution of $|p_{T,1} - p_{T,2}|$ in dijets at GEN level, with the absolute value taken for plotting convenience. 
Note that the labels $1$ and $2$ do \emph{not} correspond to leading or subleading, but rather are randomly assigned, so that the distribution $L(p_{T,1} - p_{T,2})$ is symmetric and that we apply the same initial calibration to both jets agnostic to whether it was the leading or subleading jet.
We emphasize that this is crucial to avoid introducing bias.
We show both the distributions binned in the leading SIM $p_T$ value (blue through red) and the total inclusive distribution (black).
For each distribution, we fit an exponential (\Fig{Lambda_exponential}) and a Gaussian (\Fig{Lambda_gaussian}) to extract the momentum asymmetry parameters $\Lambda_e$ and $\Lambda$ respectively, which are shown in the respective plots. 
The fits are accomplished by minimizing the mean-squared error (MSE) loss of the distribution.
We also summarize the results of these fits in \Tab{fits}, where we show the extracted momentum asymmetry values, fractional asymmetry, and fit MSE for the exponential and Gaussian.

Both by eye and by looking at the MSE's in \Tab{fits}, we can see that the exponential is a much better fit than the Gaussian, but we will consider both approaches to highlight their similarities and differences.
The extracted $\Lambda_e$ and $\Lambda$ values range from $\sim100-150$ GeV across the different $p_T$ bins.

\subsection{Correlation-Improved Jet Calibration}

We now show how Eqs.\ \eqref{eq:mean2}-\eqref{eq:var2} can be used to improve jet energy resolutions. 
As a baseline, we use CMS-prescribed resolutions, as computed in \Reference{CMS:2016lmd}:
\begin{align}
    \frac{\hat{\sigma}_{p_T}}{\hat{p_T}} &= \sqrt{\frac{\text{sgn}(N) N^2}{\hat{p_T}^2} + \frac{S^2}{\hat{p_T}} + C^2}, \label{eq:CMS_JER}
\end{align}
where $N \approx 30$ GeV, $S \approx 0.81 \text{ GeV}^{1/2}$, and $C \approx 0.04$.
This leads to a jet energy resolution of approximately $3$-$5\%$ for jets in the $p_T$ ranges we consider.

\begin{figure*}[tpbh]
\centering
    \subfloat[]{
         \includegraphics[width=\columnwidth]{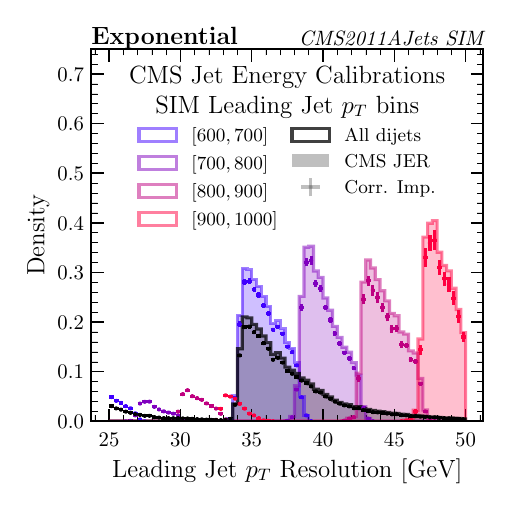}
        \label{fig:resolutions_exponential}
    }
    \subfloat[]{
        \includegraphics[width=\columnwidth]{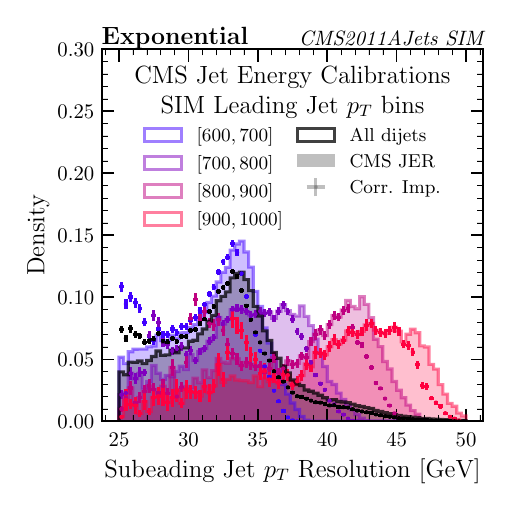}
        \label{fig:subleading_resolutions_exponential}
    }
    \caption{Distributions of (a) leading and (b) subleading jet resolutions for each $p_T$ bin, with the original CMS resolution (solid) and the correlation-improved resolutions (points), for the exponential fits.}
    \label{fig:leading_subleading_resolutions_exponential}
\end{figure*}

\begin{figure*}[tpbh]
\centering
    \subfloat[]{
         \includegraphics[width=\columnwidth]{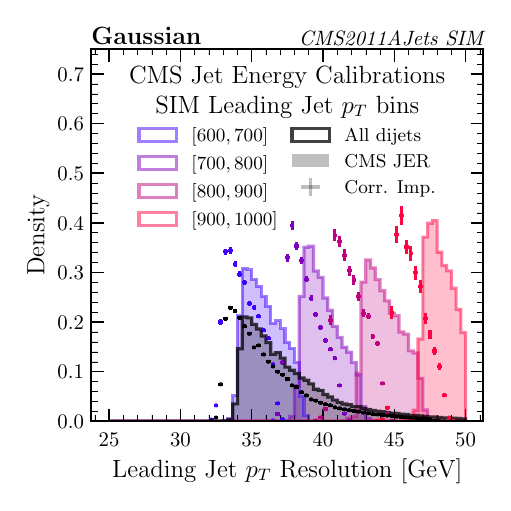}
        \label{fig:resolutions}
    }
    \subfloat[]{
        \includegraphics[width=\columnwidth]{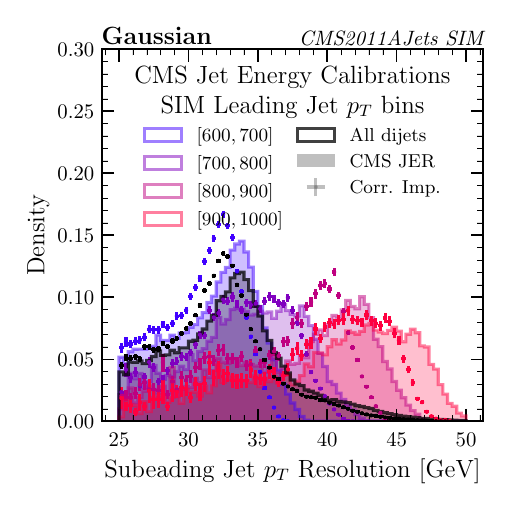}
        \label{fig:subleading_resolutions}
    }
    \caption{Distributions of (a) leading and (b) subleading jet resolutions for each $p_T$ bin, with the original CMS resolution (solid) and the correlation-improved resolutions (points), for the Gaussian fits.}
    \label{fig:leading_subleading_resolutions}
\end{figure*}

In \Figs{leading_subleading_resolutions_exponential}{leading_subleading_resolutions}, we show both the original CMS jet energy resolution (\Eq{CMS_JER}) as well as the correlation-improved resolutions (\Eqs{exponential_var2}{var2}, respectively) for both leading and subleading jets.
We show the correlation-improved resolutions for both the exponential fit in \Fig{leading_subleading_resolutions_exponential} and the Gaussian fit in \Fig{leading_subleading_resolutions}.
As with \Fig{Lambda}, events are split into bins by their leading SIM $p_T$, and we also include the inclusive distribution.
In all cases, the correlation-improved resolution distributions are shifted to the left of the original resolution distributions by a few GeV, indicating that the resolutions are indeed slightly improved. 
However, in the exponential case (\Fig{leading_subleading_resolutions_exponential}), the correlation-improved distribution is highly bimodal -- a majority of events have an unchanged resolution, whereas a small number of events (those close to the zero of the exponential, controlled by $\Lambda_e$) ``snap'' to the improved resolution.
On the other hand, in the Gaussian case, the correlation-improved distribution is a slight shift of the original distribution, reflecting the smooth behavior of the estimates in $\Lambda$.

\begin{figure*}[tpbh]
\centering
    \subfloat[]{
         \includegraphics[width=0.99\columnwidth]{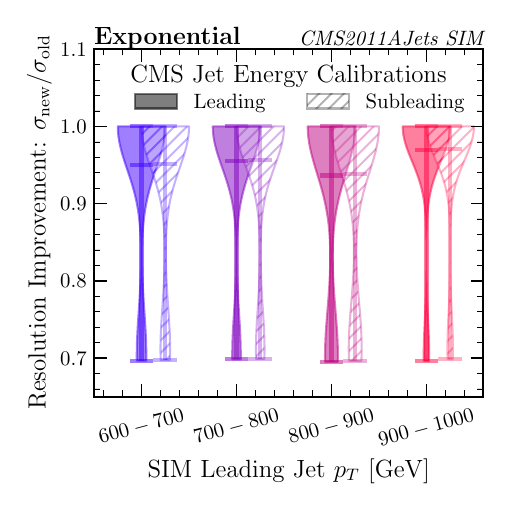}
        \label{fig:improvement_exponential}
    }
    \subfloat[]{
        \includegraphics[width=0.99\columnwidth]{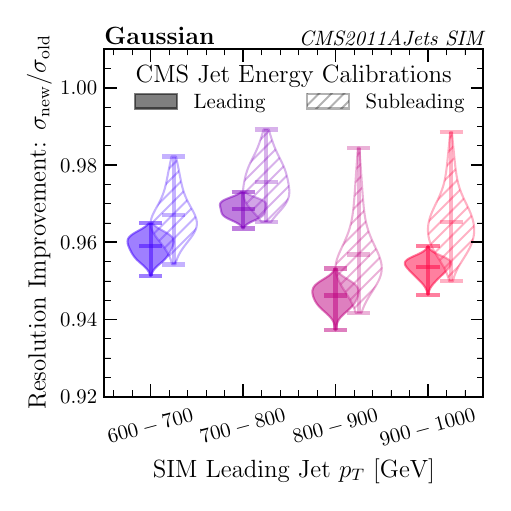}
        \label{fig:improvement_gaussian}
    }
    \caption{Violin plots showing the ratio of the correlation-improved resolution to the CMS-based resolution for each $p_T$ bin, for (a) the exponential fit and (b) the Gaussian fit. The leading jet ratio is shown with solid shading, and the subleading jet ratio is shown with lighter hatched shading. Note that the vertical axis scale is much larger for the exponential than the Gaussian, though the means of all distributions (indicated by the horizontal lines) are roughly the same at about 0.95}
    \label{fig:improvements}
\end{figure*}

Next, in \Fig{improvements}, we quantify this improvement by plotting the distribution of ratios of the new to old resolution distributions.
We can see that the resolution is improved by an average of about $3$-$5\%$ for both subleading and leading jets in the exponential and Gaussian fits, though the distribution of improvements around the average is wildly different between the two fits.
The exponential fit consists of many events with no improvement at all and few events with drastic improvement -- consistent with the bimodal picture. 
In contrast, the Gaussian fit is far more consistent, improving most events by about the same amount.
This is in accordance with the discussion in \Sec{gaussian}, where the improvement in the resolution can be approximated using \Eqs{var2}{large_lambda}for large $\Lambda$:
\begin{align}
    \frac{\sigma'_{\hat{z}_i}}{\sigma_{\hat{z}_i}} &\approx 1 - \frac{\sigma_{\hat{z}_i}^2}{2\Lambda^2},\nonumber\\
    &\approx 0.95\text{ for } \sigma_{\hat{z}_1} = \sigma_{\hat{z}_2} = 35\text{ GeV}, \Lambda = 100 \text{ GeV},\label{eq:resolution_improvement}
\end{align}
where the factor of $\frac{1}{2}$ comes from taking the square root of $\sigma^2$.

It is also common to report the quadrature improvement, which is given by:
\begin{align}
    \frac{\sqrt{\sigma'^2_{\hat{z}_i} - \sigma^2_{\hat{z}_i}}  }{\sigma_{\hat{z}_i}}.
\end{align}
For large $\Lambda$, this reduces to $\frac{\sigma_{\hat{z}_i}}{\Lambda}\approx 35\%$, which contains the same information as \Eq{resolution_improvement}

\begin{figure*}[tpbh!]
\centering
    \subfloat[]{
         \includegraphics[width=\columnwidth]{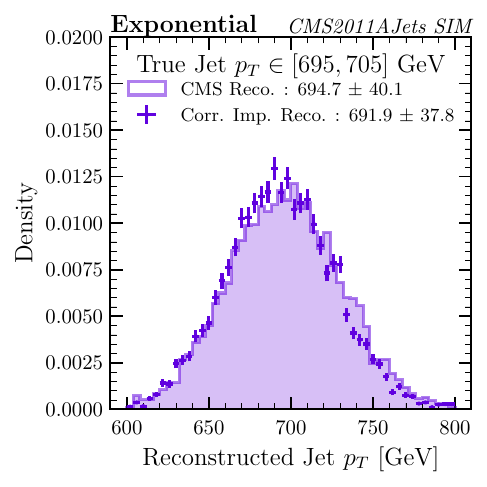}
        \label{fig:reco_exponential}
    }
    \subfloat[]{
        \includegraphics[width=\columnwidth]{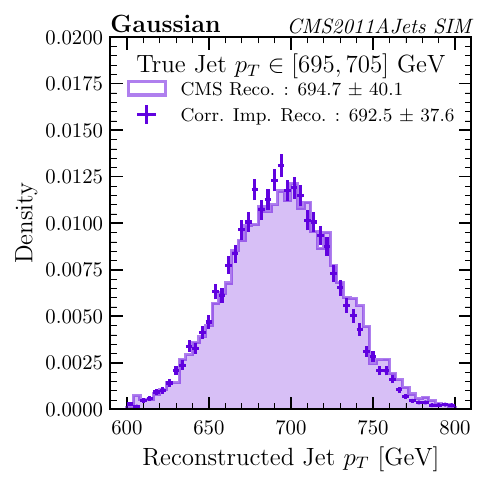}
        \label{fig:reco_gaussian}
    }
    \caption{Distribution of the reconstructed jet $p_T$ for jets with a true (GEN) $p_T \in [695, 705]$ GeV, using CMS jet energy corrections (solid, corresponding to $\hat{z}$) and the correlation-improved MLE (points, corresponding to $\hat{z}'$) for the (a) exponential and (b) Gaussian fits using \Eqs{exponential_mean}{mean2} respectively. 
    Jets are included regardless of their leading or subleading label.
    For each distribution, we write its mean and standard deviation, the latter of which should correspond to the average resolution.
    } 
    \label{fig:reco}
\end{figure*}

Finally, to check whether these estimates indeed reflect better resolutions for the $p_T$ estimators, in \Fig{reco} we plot the original CMS-estimated and correlation-improved reconstructed jet $p_T$ (corresponding to $\hat{z}$ and $\hat{z}'$, respectively) for a narrow window of true $p_T \in [695, 705]$.
For both the exponential and Gaussian fits, we see that the correlation-improved MLE does not induce any significant extra bias over the original estimates. 
We also see that the average resolution improves from $40.1$ GeV to approximately $37.5$ GeV, an improvement of approximately $6\%$ of exactly the type expected by \Eq{resolution_improvement}.
Note that the highly bimodal snapping behavior of the exponential has been washed out when looking at the full distribution of jets, and on average behaves qualitatively similar to the Gaussian.

% \begin{figure}
%     \centering
%     \includegraphics[width=0.45\textwidth]{figures/improvement.pdf}
%     \caption{The ratio of the correlation-improved resolution to the CMS-based resolution for each $p_T$ bin, for both the leading jet (dark) and subleading jet (light, hatched).}
%     \label{fig:improvement}
% \end{figure}

We end this analysis by noting that while the exponential is a better fit to the momentum asymmetry than the Gaussians in \Fig{Lambda}, both fits lead to consistent improved calibrations, as validated by \Fig{reco}.
Moreover, both methods give similar final average resolutions, in spite of the very different way $z'$ is calculated. 
As discussed in \Sec{exponential}, this is expected; both fits give the same average resolution improvement for $\Lambda \approx \Lambda_e \gg \sigma$.

\section{Discussion: Boosted Resonances}\label{sec:example_top}

The above discussions all revolved around the fact that the momenta of the two jets add to zero up to noise.
However, this is only true in the center-of-mass (COM) frame of the jets, which we always assumed we were in.
If the frame is known, one can always boost to the center-of-mass frame, where the same analysis can be performed.
However, many analyses involve the production of dijets from a boosted narrow resonance, such as a top, a Higgs boson, a $Z$, a $W^\pm$, or a new particle. 
In this case, the sum of the two jet momenta is no longer zero, but rather the momentum of the resonance, which is largely unconstrained.

If we know that the two jets originated from a boosted source of mass $M$\footnote{This is possible if the resonances as produced in pairs and the partner is independently tagged, e.g. a semileptonic $t\Bar{t}$ decay.}, then it is still possible to partially constrain the momenta of the two jets.
Rather than summing to zero, the components of the two momenta are constrained such that:
\begin{align}
    M^2 &= m_1^2 + m_2^2 \nonumber\\&
    + 2p_{T,1}p_{T,2}\left(\cosh(\eta_1 - \eta_2) - \cos(\phi_1 - \phi_2)\right) \label{eq:dijet_mass}.
\end{align}

This information can be used to improve the calibration on the $p_T$ and $m$ of both jets, assuming $\eta$ and $\phi$ are well-known.
For simplicity, suppose that we are only interested in improving just the $p_T$ calibration, and absorb the original resolution of $m_1^2$ and $m_2^2$ into $M^2$.
Define, for convenience, the quantity:
\begin{align}
    \mu \equiv \frac{M^2 - m_1^2 - m_2^2}{2\left(\cosh(\eta_1 - \eta_2) - \cos(\phi_1 - \phi_2)\right)} \label{eq:mu_definition}.
\end{align}
which is approximately the mass squared over two, for heavy resonances.
Then, the product $p_{T,1}p_{T,2}$ is approximately given by:
\begin{align}
    L(p_{T,1}p_{T,2}) = \log(\mathcal{N}(\mu, \Lambda_{\mu}^2)),
\end{align}
where $\Lambda_{\mu}$ is the resolution on $\mu$, including any effects due to resolutions on $M^2$, $m_{1,2}^2$, $\phi_{1,2}$, and $\eta_{1,2}$.
Note that unlike the cases considered previously, this is a constraint on the \emph{product} of the $p_T$'s, not the sum of the $p_T$'s.
It is also possible, as before, to consider exponentials rather than Gaussians here.

\begin{figure}[tpb]
    \centering
    \includegraphics[width=\columnwidth]{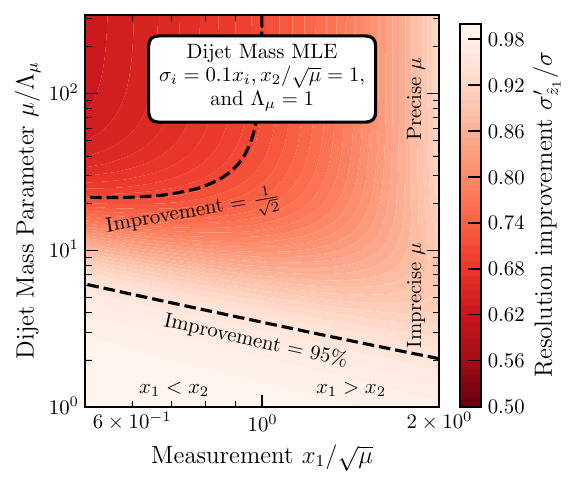}
    \caption{Plot showing the ratio of the correlation-improved resolution to the original resolution for one of the two jets, as a function of the measured $p_T$, $x_1/\sqrt{\mu}$, and the dijet mass parameter $\mu/\Lambda_\mu$ as defined in \Eq{mu_definition}. We fix the original jet energy resolutions $\sigma_{\hat{z}_1}$ and $\sigma_{\hat{z}_1}$ to be $10\%$ of the nominal energy, and we also fix $x_2/\sqrt{\mu} = 1$ and $\Lambda_\mu$ for simplicity.}
    \label{fig:dijet_mass}
\end{figure}

We can now  take derivatives of \Eq{general_likelihood_z} to solve for the correlation-improved estimates and resolutions for $z_1 = p_{T,1}$ and $z_2 = p_{T,2}$. 
In general, this requires solving a system of coupled nonlinear equations and lacks a closed-form solution.
We perform this numerically, and present the resolution improvement, $\sigma'_{\hat{z}_1} / \sigma_{\hat{z}_1}$ in \Fig{dijet_mass} for example parameter choices.
We see that the resolution indeed improves -- the improvement is more pronounced if $\mu/\Lambda_\mu$ is small, with the typical $1/\sqrt{2}$ improvement factor that occurs whenever $\sigma_{\hat{z}_1} \sim \sigma_{\hat{z}_2} \gg \Lambda_\mu$. 
Note that the results in \Fig{dijet_mass} are just purely numeric for example parameter choices, and there are no jets (real or simulated) here.
We leave the study of these results in the setting of actual or simulated jets to potential future work.

% \subsection{Dijet Cross Sections}

% Application to jet $pT$ cross section measurement

% Unlike the jet $p_T$ or the individual jet masses, the dijet mass does \emph{not} improve

As an aside, constraining the jet $p_T$ using momentum conservation as we have done in \Sec{example_qcd} does \emph{not} help to improve the dijet resolution. 
This is because the dijet mass is calculated using the sum of the jet momenta, which already contains all of the information about momentum conservation.

\section{Conclusion}\label{sec:conclusion}

In this paper, we have shown how one can take advantage of correlations between jets, namely those induced by momentum conservation in dijet events, to obtain jet-by-jet correlation-improved maximum likelihood estimates and resolutions.
Using this, we are able to achieve improvements in resolution over the baseline of $3$-$5\%$ (35\% quadrature improvement) for simulated dijet events from the CMS detector.
This is possible by only making assumptions on the relative momenta between the jets, without making assumptions on the overall prior of jet production, and does not introduce any bias as long as the jets remain unordered.
This is in contrast to other proposals for improving jet energy calibrations, which focus on the high-dimensional substructure and detector interactions of individual jets, or proposals that use momentum conservation to constrain the statistical properties of jets. %global properties of the event.

While we have primarily focused on the application of correlation-improved MLE to dijet events in the COM frame, the procedure of applying \Eqs{MLE_mean}{MLE_std} to \Eq{general_likelihood_z} can be expanded.
For instance, we have discussed briefly how one can loosen the constraint of momentum conservation in the COM frame to simply knowing the dijet mass in \Sec{example_top}.
Other types of correlations between jets, such as those induced by color connections~\cite{Gallicchio:2010sw, D0:2011lzz, ATLAS:2015ytt, ATLAS:2018olo, Cavallini:2021vot}, can also be used here.
Additionally, we have used CMS-prescribed jet energy resolutions and calibrations as our baseline, though in principle one could use any (especially machine learned) unbiased calibration for $L(x_i|z_i)$ to source $\hat{z}(x)$ and $\sigma_{\hat{z}}(x)$ for the correlation-improved MLE, such as the Gaussian Ansatz of \Reference{Gambhir:2022gua}.
Since we are only using out-of-jet information, correlation-improved MLE can be used on top of any type of single-jet calibration method.

An important aspect of this technique is fitting the distribution of the momentum asymmetry $z_1 - z_2$ at truth level.
This requires that the simulation of the relevant effects causing the momentum asymmetry is reliable.
However, because of the way prior dependence enters \Eq{general_likelihood_z}, \emph{only} the momentum asymmetry needs to be reliable -- the distribution of $z_1$ or $z_2$ themselves are not important, which is a desirable feature of calibration. 
Moreover, as we have shown with the Gaussian fit, it is possible to still get a consistent and valid improvement, even if the fit is sub-optimal.
Global event features are therefore a robust way to add calibration information, and in the future these features can offer an approach complementary to modern ML methods to continue improving jet energy calibrations.

\section*{Code and Data}

The CMS Open Data used in \Sec{example_qcd} was accessed in the MIT Open Data (MOD) format and can be found at \url{https://energyflow.network/docs/datasets/}~\cite{komiske_patrick_2019_3341502,komiske_patrick_2019_3341770,komiske_patrick_2019_3341772,nachman_benjamin_2021_5108967}.
The code used for the analyses presented in this paper is located at \url{https://github.com/rikab/SeeingDouble}, from which all of the results and plots shown here may be reproduced.

\vspace{5mm}

\begin{acknowledgments}

We would like to thank Aviv Cukierman for useful discussions, as well as Sean Benevedes and Jesse Thaler for feedback on the manuscript.
BN is supported by the U.S. Department of Energy (DOE), Office of Science under contract DE-AC02-05CH11231.
RG is supported by the National Science Foundation under Cooperative Agreement PHY-2019786 (The NSF AI Institute for Artificial Intelligence and Fundamental Interactions, \url{http://iaifi.org/}), and by the U.S. DOE Office of High Energy Physics under grant number DE-SC0012567.
RG thanks the Galileo Galilei Institute for Theoretical Physics for their hospitality and the INFN for support during the completion of this work.
\end{acknowledgments}

\bibliography{myrefs,HEPML}

\begin{thebibliography}{56}%
\makeatletter
\providecommand \@ifxundefined [1]{%
 \@ifx{#1\undefined}
}%
\providecommand \@ifnum [1]{%
 \ifnum #1\expandafter \@firstoftwo
 \else \expandafter \@secondoftwo
 \fi
}%
\providecommand \@ifx [1]{%
 \ifx #1\expandafter \@firstoftwo
 \else \expandafter \@secondoftwo
 \fi
}%
\providecommand \natexlab [1]{#1}%
\providecommand \enquote  [1]{``#1''}%
\providecommand \bibnamefont  [1]{#1}%
\providecommand \bibfnamefont [1]{#1}%
\providecommand \citenamefont [1]{#1}%
\providecommand \href@noop [0]{\@secondoftwo}%
\providecommand \href [0]{\begingroup \@sanitize@url \@href}%
\providecommand \@href[1]{\@@startlink{#1}\@@href}%
\providecommand \@@href[1]{\endgroup#1\@@endlink}%
\providecommand \@sanitize@url [0]{\catcode `\\12\catcode `\$12\catcode
  `\&12\catcode `\#12\catcode `\^12\catcode `\_12\catcode `\%12\relax}%
\providecommand \@@startlink[1]{}%
\providecommand \@@endlink[0]{}%
\providecommand \url  [0]{\begingroup\@sanitize@url \@url }%
\providecommand \@url [1]{\endgroup\@href {#1}{\urlprefix }}%
\providecommand \urlprefix  [0]{URL }%
\providecommand \Eprint [0]{\href }%
\providecommand \doibase [0]{http://dx.doi.org/}%
\providecommand \selectlanguage [0]{\@gobble}%
\providecommand \bibinfo  [0]{\@secondoftwo}%
\providecommand \bibfield  [0]{\@secondoftwo}%
\providecommand \translation [1]{[#1]}%
\providecommand \BibitemOpen [0]{}%
\providecommand \bibitemStop [0]{}%
\providecommand \bibitemNoStop [0]{.\EOS\space}%
\providecommand \EOS [0]{\spacefactor3000\relax}%
\providecommand \BibitemShut  [1]{\csname bibitem#1\endcsname}%
\let\auto@bib@innerbib\@empty
%</preamble>
\bibitem [{\citenamefont {collaboration}(2011)}]{collaboration_2011}%
  \BibitemOpen
  \bibfield  {author} {\bibinfo {author} {\bibfnamefont {The~CMS}\ \bibnamefont
  {collaboration}},\ }\bibfield  {title} {\enquote {\bibinfo {title}
  {Determination of jet energy calibration and transverse momentum resolution
  in cms},}\ }\href {\doibase 10.1088/1748-0221/6/11/p11002} {\bibfield
  {journal} {\bibinfo  {journal} {Journal of Instrumentation}\ }\textbf
  {\bibinfo {volume} {6}},\ \bibinfo {pages} {P11002–P11002} (\bibinfo {year}
  {2011})}\BibitemShut {NoStop}%
\bibitem [{\citenamefont {Khachatryan}\ \emph {et~al.}(2017)\citenamefont
  {Khachatryan} \emph {et~al.}}]{CMS:2016lmd}%
  \BibitemOpen
  \bibfield  {author} {\bibinfo {author} {\bibfnamefont {Vardan}\ \bibnamefont
  {Khachatryan}} \emph {et~al.} (\bibinfo {collaboration} {CMS}),\ }\bibfield
  {title} {\enquote {\bibinfo {title} {{Jet energy scale and resolution in the
  CMS experiment in pp collisions at 8 TeV}},}\ }\href {\doibase
  10.1088/1748-0221/12/02/P02014} {\bibfield  {journal} {\bibinfo  {journal}
  {JINST}\ }\textbf {\bibinfo {volume} {12}},\ \bibinfo {pages} {P02014}
  (\bibinfo {year} {2017})},\ \Eprint {http://arxiv.org/abs/1607.03663}
  {arXiv:1607.03663 [hep-ex]} \BibitemShut {NoStop}%
\bibitem [{\citenamefont {Aaboud}\ \emph {et~al.}(2020)\citenamefont {Aaboud}
  \emph {et~al.}}]{ATLAS:2019oxp}%
  \BibitemOpen
  \bibfield  {author} {\bibinfo {author} {\bibfnamefont {Morad}\ \bibnamefont
  {Aaboud}} \emph {et~al.} (\bibinfo {collaboration} {ATLAS}),\ }\bibfield
  {title} {\enquote {\bibinfo {title} {{Determination of jet calibration and
  energy resolution in proton-proton collisions at $\sqrt{s}$ = 8 TeV using the
  ATLAS detector}},}\ }\href {\doibase 10.1140/epjc/s10052-020-08477-8}
  {\bibfield  {journal} {\bibinfo  {journal} {Eur. Phys. J. C}\ }\textbf
  {\bibinfo {volume} {80}},\ \bibinfo {pages} {1104} (\bibinfo {year}
  {2020})},\ \Eprint {http://arxiv.org/abs/1910.04482} {arXiv:1910.04482
  [hep-ex]} \BibitemShut {NoStop}%
\bibitem [{\citenamefont {Aad}\ \emph {et~al.}(2021)\citenamefont {Aad} \emph
  {et~al.}}]{ATLAS:2020cli}%
  \BibitemOpen
  \bibfield  {author} {\bibinfo {author} {\bibfnamefont {Georges}\ \bibnamefont
  {Aad}} \emph {et~al.} (\bibinfo {collaboration} {ATLAS}),\ }\bibfield
  {title} {\enquote {\bibinfo {title} {{Jet energy scale and resolution
  measured in proton\textendash{}proton collisions at $\sqrt{s}=13$~TeV with
  the ATLAS detector}},}\ }\href {\doibase 10.1140/epjc/s10052-021-09402-3}
  {\bibfield  {journal} {\bibinfo  {journal} {Eur. Phys. J. C}\ }\textbf
  {\bibinfo {volume} {81}},\ \bibinfo {pages} {689} (\bibinfo {year} {2021})},\
  \Eprint {http://arxiv.org/abs/2007.02645} {arXiv:2007.02645 [hep-ex]}
  \BibitemShut {NoStop}%
\bibitem [{CMS(2020)}]{CMS-DP-2020-019}%
  \BibitemOpen
  \bibfield  {title} {\enquote {\bibinfo {title} {{Jet energy scale and
  resolution performance with 13 TeV data collected by CMS in 2016-2018}},}\
  }\href {https://cds.cern.ch/record/2715872} {\  (\bibinfo {year}
  {2020})}\BibitemShut {NoStop}%
\bibitem [{\citenamefont {Aad}\ \emph {et~al.}(2023)\citenamefont {Aad} \emph
  {et~al.}}]{ATLAS:2023tyv}%
  \BibitemOpen
  \bibfield  {author} {\bibinfo {author} {\bibfnamefont {Georges}\ \bibnamefont
  {Aad}} \emph {et~al.} (\bibinfo {collaboration} {ATLAS}),\ }\bibfield
  {title} {\enquote {\bibinfo {title} {{New techniques for jet calibration with
  the ATLAS detector}},}\ }\href {\doibase 10.1140/epjc/s10052-023-11837-9}
  {\bibfield  {journal} {\bibinfo  {journal} {Eur. Phys. J. C}\ }\textbf
  {\bibinfo {volume} {83}},\ \bibinfo {pages} {761} (\bibinfo {year} {2023})},\
  \Eprint {http://arxiv.org/abs/2303.17312} {arXiv:2303.17312 [hep-ex]}
  \BibitemShut {NoStop}%
\bibitem [{\citenamefont {Komiske}\ \emph {et~al.}(2017)\citenamefont
  {Komiske}, \citenamefont {Metodiev}, \citenamefont {Nachman},\ and\
  \citenamefont {Schwartz}}]{Komiske:2017ubm}%
  \BibitemOpen
  \bibfield  {author} {\bibinfo {author} {\bibfnamefont {Patrick~T.}\
  \bibnamefont {Komiske}}, \bibinfo {author} {\bibfnamefont {Eric~M.}\
  \bibnamefont {Metodiev}}, \bibinfo {author} {\bibfnamefont {Benjamin}\
  \bibnamefont {Nachman}}, \ and\ \bibinfo {author} {\bibfnamefont
  {Matthew~D.}\ \bibnamefont {Schwartz}},\ }\bibfield  {title} {\enquote
  {\bibinfo {title} {{Pileup Mitigation with Machine Learning (PUMML)}},}\
  }\href {\doibase 10.1007/JHEP12(2017)051} {\bibfield  {journal} {\bibinfo
  {journal} {JHEP}\ }\textbf {\bibinfo {volume} {12}},\ \bibinfo {pages} {051}
  (\bibinfo {year} {2017})},\ \Eprint {http://arxiv.org/abs/1707.08600}
  {arXiv:1707.08600 [hep-ph]} \BibitemShut {NoStop}%
\bibitem [{\citenamefont {Arjona~Mart\'\i{}nez}\ \emph
  {et~al.}(2019)\citenamefont {Arjona~Mart\'\i{}nez}, \citenamefont {Cerri},
  \citenamefont {Pierini}, \citenamefont {Spiropulu},\ and\ \citenamefont
  {Vlimant}}]{ArjonaMartinez:2018eah}%
  \BibitemOpen
  \bibfield  {author} {\bibinfo {author} {\bibfnamefont {J.}~\bibnamefont
  {Arjona~Mart\'\i{}nez}}, \bibinfo {author} {\bibfnamefont {Olmo}\
  \bibnamefont {Cerri}}, \bibinfo {author} {\bibfnamefont {Maurizio}\
  \bibnamefont {Pierini}}, \bibinfo {author} {\bibfnamefont {Maria}\
  \bibnamefont {Spiropulu}}, \ and\ \bibinfo {author} {\bibfnamefont
  {Jean-Roch}\ \bibnamefont {Vlimant}},\ }\bibfield  {title} {\enquote
  {\bibinfo {title} {{Pileup mitigation at the Large Hadron Collider with graph
  neural networks}},}\ }\href {\doibase 10.1140/epjp/i2019-12710-3} {\bibfield
  {journal} {\bibinfo  {journal} {Eur. Phys. J. Plus}\ }\textbf {\bibinfo
  {volume} {134}},\ \bibinfo {pages} {333} (\bibinfo {year} {2019})},\ \Eprint
  {http://arxiv.org/abs/1810.07988} {arXiv:1810.07988 [hep-ph]} \BibitemShut
  {NoStop}%
\bibitem [{\citenamefont {{ATLAS
  Collaboration}}(2018)}]{ATL-PHYS-PUB-2018-013}%
  \BibitemOpen
  \bibfield  {author} {\bibinfo {author} {\bibnamefont {{ATLAS
  Collaboration}}},\ }\bibfield  {title} {\enquote {\bibinfo {title}
  {{Generalized Numerical Inversion: A Neural Network Approach to Jet
  Calibration}},}\ }\href {http://cdsweb.cern.ch/record/2630972} {\bibfield
  {journal} {\bibinfo  {journal} {ATL-PHYS-PUB-2018-013}\ } (\bibinfo {year}
  {2018})}\BibitemShut {NoStop}%
\bibitem [{\citenamefont {Arjona~Martínez}\ \emph {et~al.}(2019)\citenamefont
  {Arjona~Martínez}, \citenamefont {Cerri}, \citenamefont {Pierini},
  \citenamefont {Spiropulu},\ and\ \citenamefont {Vlimant}}]{Martinez:2018fwc}%
  \BibitemOpen
  \bibfield  {author} {\bibinfo {author} {\bibfnamefont {J.}~\bibnamefont
  {Arjona~Martínez}}, \bibinfo {author} {\bibfnamefont {Olmo}\ \bibnamefont
  {Cerri}}, \bibinfo {author} {\bibfnamefont {Maurizio}\ \bibnamefont
  {Pierini}}, \bibinfo {author} {\bibfnamefont {Maria}\ \bibnamefont
  {Spiropulu}}, \ and\ \bibinfo {author} {\bibfnamefont {Jean-Roch}\
  \bibnamefont {Vlimant}},\ }\bibfield  {title} {\enquote {\bibinfo {title}
  {{Pileup mitigation at the Large Hadron Collider with graph neural
  networks}},}\ }\href {\doibase 10.1140/epjp/i2019-12710-3} {\bibfield
  {journal} {\bibinfo  {journal} {Eur. Phys. J. Plus}\ }\textbf {\bibinfo
  {volume} {134}},\ \bibinfo {pages} {333} (\bibinfo {year} {2019})},\ \Eprint
  {http://arxiv.org/abs/1810.07988} {arXiv:1810.07988 [hep-ph]} \BibitemShut
  {NoStop}%
\bibitem [{\citenamefont {Haake}\ and\ \citenamefont
  {Loizides}(2019)}]{Haake:2018hqn}%
  \BibitemOpen
  \bibfield  {author} {\bibinfo {author} {\bibfnamefont {R\"udiger}\
  \bibnamefont {Haake}}\ and\ \bibinfo {author} {\bibfnamefont {Constantin}\
  \bibnamefont {Loizides}},\ }\bibfield  {title} {\enquote {\bibinfo {title}
  {{Machine Learning based jet momentum reconstruction in heavy-ion
  collisions}},}\ }\href {\doibase 10.1103/PhysRevC.99.064904} {\bibfield
  {journal} {\bibinfo  {journal} {Phys. Rev. C}\ }\textbf {\bibinfo {volume}
  {99}},\ \bibinfo {pages} {064904} (\bibinfo {year} {2019})},\ \Eprint
  {http://arxiv.org/abs/1810.06324} {arXiv:1810.06324 [nucl-ex]} \BibitemShut
  {NoStop}%
\bibitem [{\citenamefont {Haake}(2020)}]{Haake:2019pqd}%
  \BibitemOpen
  \bibfield  {author} {\bibinfo {author} {\bibfnamefont {R\"udiger}\
  \bibnamefont {Haake}} (\bibinfo {collaboration} {ALICE}),\ }\bibfield
  {title} {\enquote {\bibinfo {title} {{Machine Learning based jet momentum
  reconstruction in Pb-Pb collisions measured with the ALICE detector}},}\
  }\href {\doibase 10.22323/1.364.0312} {\bibfield  {journal} {\bibinfo
  {journal} {PoS}\ }\textbf {\bibinfo {volume} {EPS-HEP2019}},\ \bibinfo
  {pages} {312} (\bibinfo {year} {2020})},\ \Eprint
  {http://arxiv.org/abs/1909.01639} {arXiv:1909.01639 [nucl-ex]} \BibitemShut
  {NoStop}%
\bibitem [{\citenamefont {Sirunyan}\ \emph {et~al.}(2020)\citenamefont
  {Sirunyan} \emph {et~al.}}]{CMS:2019uxx}%
  \BibitemOpen
  \bibfield  {author} {\bibinfo {author} {\bibfnamefont {Albert~M}\
  \bibnamefont {Sirunyan}} \emph {et~al.} (\bibinfo {collaboration} {CMS}),\
  }\bibfield  {title} {\enquote {\bibinfo {title} {{A Deep Neural Network for
  Simultaneous Estimation of b Jet Energy and Resolution}},}\ }\href {\doibase
  10.1007/s41781-020-00041-z} {\bibfield  {journal} {\bibinfo  {journal}
  {Comput. Softw. Big Sci.}\ }\textbf {\bibinfo {volume} {4}},\ \bibinfo
  {pages} {10} (\bibinfo {year} {2020})},\ \Eprint
  {http://arxiv.org/abs/1912.06046} {arXiv:1912.06046 [hep-ex]} \BibitemShut
  {NoStop}%
\bibitem [{ATL(2019)}]{ATL-PHYS-PUB-2019-028}%
  \BibitemOpen
  \href {https://cds.cern.ch/record/2684070} {\emph {\bibinfo {title}
  {{Convolutional Neural Networks with Event Images for Pileup Mitigation with
  the ATLAS Detector}}}},\ \bibinfo {type} {Tech. Rep.}\ (\bibinfo
  {institution} {CERN},\ \bibinfo {address} {Geneva},\ \bibinfo {year}
  {2019})\BibitemShut {NoStop}%
\bibitem [{\citenamefont {Carrazza}\ and\ \citenamefont
  {Dreyer}(2019)}]{Carrazza:2019efs}%
  \BibitemOpen
  \bibfield  {author} {\bibinfo {author} {\bibfnamefont {Stefano}\ \bibnamefont
  {Carrazza}}\ and\ \bibinfo {author} {\bibfnamefont {Frédéric~A.}\
  \bibnamefont {Dreyer}},\ }\bibfield  {title} {\enquote {\bibinfo {title}
  {{Jet grooming through reinforcement learning}},}\ }\href {\doibase
  10.1103/PhysRevD.100.014014} {\bibfield  {journal} {\bibinfo  {journal}
  {Phys. Rev. D}\ }\textbf {\bibinfo {volume} {100}},\ \bibinfo {pages}
  {014014} (\bibinfo {year} {2019})},\ \Eprint
  {http://arxiv.org/abs/1903.09644} {arXiv:1903.09644 [hep-ph]} \BibitemShut
  {NoStop}%
\bibitem [{\citenamefont {{ATLAS
  Collaboration}}(2020)}]{ATL-PHYS-PUB-2020-001}%
  \BibitemOpen
  \bibfield  {author} {\bibinfo {author} {\bibnamefont {{ATLAS
  Collaboration}}},\ }\bibfield  {title} {\enquote {\bibinfo {title}
  {{Simultaneous Jet Energy and Mass Calibrations with Neural Networks}},}\
  }\href {http://cdsweb.cern.ch/record/2706189} {\bibfield  {journal} {\bibinfo
   {journal} {ATL-PHYS-PUB-2020-001}\ } (\bibinfo {year} {2020})}\BibitemShut
  {NoStop}%
\bibitem [{\citenamefont {{Baldi, Pierre and Blecher, Lukas and Butter, Anja
  and Collado, Julian and Howard, Jessica N. and Keilbach, Fabian and Plehn,
  Tilman and Kasieczka, Gregor and Whiteson, Daniel}}(2020)}]{Baldi:2020hjm}%
  \BibitemOpen
  \bibfield  {author} {\bibinfo {author} {\bibnamefont {{Baldi, Pierre and
  Blecher, Lukas and Butter, Anja and Collado, Julian and Howard, Jessica N.
  and Keilbach, Fabian and Plehn, Tilman and Kasieczka, Gregor and Whiteson,
  Daniel}}},\ }\bibfield  {title} {\enquote {\bibinfo {title} {{How to GAN
  Higher Jet Resolution}},}\ }\href@noop {} {\  (\bibinfo {year} {2020})},\
  \Eprint {http://arxiv.org/abs/2012.11944} {arXiv:2012.11944 [hep-ph]}
  \BibitemShut {NoStop}%
\bibitem [{\citenamefont {Kasieczka}\ \emph {et~al.}(2020)\citenamefont
  {Kasieczka}, \citenamefont {Luchmann}, \citenamefont {Otterpohl},\ and\
  \citenamefont {Plehn}}]{Kasieczka:2020vlh}%
  \BibitemOpen
  \bibfield  {author} {\bibinfo {author} {\bibfnamefont {Gregor}\ \bibnamefont
  {Kasieczka}}, \bibinfo {author} {\bibfnamefont {Michel}\ \bibnamefont
  {Luchmann}}, \bibinfo {author} {\bibfnamefont {Florian}\ \bibnamefont
  {Otterpohl}}, \ and\ \bibinfo {author} {\bibfnamefont {Tilman}\ \bibnamefont
  {Plehn}},\ }\href {\doibase 10.21468/SciPostPhys.9.6.089} {\enquote {\bibinfo
  {title} {{Per-Object Systematics using Deep-Learned Calibration}},}\ }
  (\bibinfo {year} {2020}),\ \Eprint {http://arxiv.org/abs/2003.11099}
  {arXiv:2003.11099 [hep-ph]} \BibitemShut {NoStop}%
\bibitem [{\citenamefont {Maier}\ \emph {et~al.}(2021)\citenamefont {Maier},
  \citenamefont {Narayanan}, \citenamefont {de~Castro}, \citenamefont
  {Goncharov}, \citenamefont {Paus},\ and\ \citenamefont
  {Schott}}]{Maier:2021ymx}%
  \BibitemOpen
  \bibfield  {author} {\bibinfo {author} {\bibfnamefont {Benedikt}\
  \bibnamefont {Maier}}, \bibinfo {author} {\bibfnamefont {Siddharth~M.}\
  \bibnamefont {Narayanan}}, \bibinfo {author} {\bibfnamefont {Gianfranco}\
  \bibnamefont {de~Castro}}, \bibinfo {author} {\bibfnamefont {Maxim}\
  \bibnamefont {Goncharov}}, \bibinfo {author} {\bibfnamefont {Christoph}\
  \bibnamefont {Paus}}, \ and\ \bibinfo {author} {\bibfnamefont {Matthias}\
  \bibnamefont {Schott}},\ }\bibfield  {title} {\enquote {\bibinfo {title}
  {{Pile-Up Mitigation using Attention}},}\ }\href {\doibase
  10.1088/2632-2153/ac7198} {\bibfield  {journal} {\bibinfo  {journal}
  {Mach.Learn.Sci.Tech.}\ }\textbf {\bibinfo {volume} {3}},\ \bibinfo {pages}
  {025012} (\bibinfo {year} {2021})},\ \Eprint
  {http://arxiv.org/abs/2107.02779} {arXiv:2107.02779 [physics.ins-det]}
  \BibitemShut {NoStop}%
\bibitem [{\citenamefont {Gambhir}\ \emph
  {et~al.}(2022{\natexlab{a}})\citenamefont {Gambhir}, \citenamefont
  {Nachman},\ and\ \citenamefont {Thaler}}]{Gambhir:2022gua}%
  \BibitemOpen
  \bibfield  {author} {\bibinfo {author} {\bibfnamefont {Rikab}\ \bibnamefont
  {Gambhir}}, \bibinfo {author} {\bibfnamefont {Benjamin}\ \bibnamefont
  {Nachman}}, \ and\ \bibinfo {author} {\bibfnamefont {Jesse}\ \bibnamefont
  {Thaler}},\ }\bibfield  {title} {\enquote {\bibinfo {title} {{Learning
  Uncertainties the Frequentist Way: Calibration and Correlation in High Energy
  Physics}},}\ }\href {\doibase 10.1103/PhysRevLett.129.082001} {\bibfield
  {journal} {\bibinfo  {journal} {Phys. Rev. Lett.}\ }\textbf {\bibinfo
  {volume} {129}},\ \bibinfo {pages} {082001} (\bibinfo {year}
  {2022}{\natexlab{a}})},\ \Eprint {http://arxiv.org/abs/2205.03413}
  {arXiv:2205.03413 [hep-ph]} \BibitemShut {NoStop}%
\bibitem [{\citenamefont {Gambhir}\ \emph
  {et~al.}(2022{\natexlab{b}})\citenamefont {Gambhir}, \citenamefont
  {Nachman},\ and\ \citenamefont {Thaler}}]{Gambhir:2022dut}%
  \BibitemOpen
  \bibfield  {author} {\bibinfo {author} {\bibfnamefont {Rikab}\ \bibnamefont
  {Gambhir}}, \bibinfo {author} {\bibfnamefont {Benjamin}\ \bibnamefont
  {Nachman}}, \ and\ \bibinfo {author} {\bibfnamefont {Jesse}\ \bibnamefont
  {Thaler}},\ }\bibfield  {title} {\enquote {\bibinfo {title} {{Bias and priors
  in machine learning calibrations for high energy physics}},}\ }\href
  {\doibase 10.1103/PhysRevD.106.036011} {\bibfield  {journal} {\bibinfo
  {journal} {Phys. Rev. D}\ }\textbf {\bibinfo {volume} {106}},\ \bibinfo
  {pages} {036011} (\bibinfo {year} {2022}{\natexlab{b}})},\ \Eprint
  {http://arxiv.org/abs/2205.05084} {arXiv:2205.05084 [hep-ph]} \BibitemShut
  {NoStop}%
\bibitem [{\citenamefont {Li}\ \emph {et~al.}(2023)\citenamefont {Li},
  \citenamefont {Liu}, \citenamefont {Feng}, \citenamefont {Paspalaki},
  \citenamefont {Tran}, \citenamefont {Liu},\ and\ \citenamefont
  {Li}}]{Li:2022omf}%
  \BibitemOpen
  \bibfield  {author} {\bibinfo {author} {\bibfnamefont {Tianchun}\
  \bibnamefont {Li}}, \bibinfo {author} {\bibfnamefont {Shikun}\ \bibnamefont
  {Liu}}, \bibinfo {author} {\bibfnamefont {Yongbin}\ \bibnamefont {Feng}},
  \bibinfo {author} {\bibfnamefont {Garyfallia}\ \bibnamefont {Paspalaki}},
  \bibinfo {author} {\bibfnamefont {Nhan~V.}\ \bibnamefont {Tran}}, \bibinfo
  {author} {\bibfnamefont {Miaoyuan}\ \bibnamefont {Liu}}, \ and\ \bibinfo
  {author} {\bibfnamefont {Pan}\ \bibnamefont {Li}},\ }\bibfield  {title}
  {\enquote {\bibinfo {title} {{Semi-supervised graph neural networks for
  pileup noise removal}},}\ }\href {\doibase 10.1140/epjc/s10052-022-11083-5}
  {\bibfield  {journal} {\bibinfo  {journal} {Eur. Phys. J. C}\ }\textbf
  {\bibinfo {volume} {83}},\ \bibinfo {pages} {99} (\bibinfo {year} {2023})},\
  \Eprint {http://arxiv.org/abs/2203.15823} {arXiv:2203.15823 [hep-ex]}
  \BibitemShut {NoStop}%
\bibitem [{\citenamefont {Angloher}\ \emph {et~al.}(2023)\citenamefont
  {Angloher} \emph {et~al.}}]{CRESST:2022qor}%
  \BibitemOpen
  \bibfield  {author} {\bibinfo {author} {\bibfnamefont {G.}~\bibnamefont
  {Angloher}} \emph {et~al.} (\bibinfo {collaboration} {CRESST}),\ }\bibfield
  {title} {\enquote {\bibinfo {title} {{Towards an automated data cleaning with
  deep learning in CRESST}},}\ }\href {\doibase
  10.1140/epjp/s13360-023-03674-2} {\bibfield  {journal} {\bibinfo  {journal}
  {Eur. Phys. J. Plus}\ }\textbf {\bibinfo {volume} {138}},\ \bibinfo {pages}
  {100} (\bibinfo {year} {2023})},\ \Eprint {http://arxiv.org/abs/2211.00564}
  {arXiv:2211.00564 [physics.ins-det]} \BibitemShut {NoStop}%
\bibitem [{\citenamefont {Gouskos}\ \emph {et~al.}(2023)\citenamefont
  {Gouskos}, \citenamefont {Iemmi}, \citenamefont {Liechti}, \citenamefont
  {Maier}, \citenamefont {Mikuni},\ and\ \citenamefont {Qu}}]{Gouskos:2022gjg}%
  \BibitemOpen
  \bibfield  {author} {\bibinfo {author} {\bibfnamefont {Loukas}\ \bibnamefont
  {Gouskos}}, \bibinfo {author} {\bibfnamefont {Fabio}\ \bibnamefont {Iemmi}},
  \bibinfo {author} {\bibfnamefont {Sascha}\ \bibnamefont {Liechti}}, \bibinfo
  {author} {\bibfnamefont {Benedikt}\ \bibnamefont {Maier}}, \bibinfo {author}
  {\bibfnamefont {Vinicius}\ \bibnamefont {Mikuni}}, \ and\ \bibinfo {author}
  {\bibfnamefont {Huilin}\ \bibnamefont {Qu}},\ }\bibfield  {title} {\enquote
  {\bibinfo {title} {{Optimal transport for a novel event description at hadron
  colliders}},}\ }\href {\doibase 10.1103/PhysRevD.108.096003} {\bibfield
  {journal} {\bibinfo  {journal} {Phys. Rev. D}\ }\textbf {\bibinfo {volume}
  {108}},\ \bibinfo {pages} {096003} (\bibinfo {year} {2023})},\ \Eprint
  {http://arxiv.org/abs/2211.02029} {arXiv:2211.02029 [hep-ph]} \BibitemShut
  {NoStop}%
\bibitem [{\citenamefont {Kim}\ \emph {et~al.}(2023)\citenamefont {Kim},
  \citenamefont {Ahn}, \citenamefont {Chae}, \citenamefont {Hooker},\ and\
  \citenamefont {Rogachev}}]{Kim:2023koz}%
  \BibitemOpen
  \bibfield  {author} {\bibinfo {author} {\bibfnamefont {C.~H.}\ \bibnamefont
  {Kim}}, \bibinfo {author} {\bibfnamefont {S.}~\bibnamefont {Ahn}}, \bibinfo
  {author} {\bibfnamefont {K.~Y.}\ \bibnamefont {Chae}}, \bibinfo {author}
  {\bibfnamefont {J.}~\bibnamefont {Hooker}}, \ and\ \bibinfo {author}
  {\bibfnamefont {G.~V.}\ \bibnamefont {Rogachev}},\ }\bibfield  {title}
  {\enquote {\bibinfo {title} {{Restoring original signals from pile-up using
  deep learning}},}\ }\href {\doibase 10.1016/j.nima.2023.168492} {\bibfield
  {journal} {\bibinfo  {journal} {Nucl. Instrum. Meth. A}\ }\textbf {\bibinfo
  {volume} {1055}},\ \bibinfo {pages} {168492} (\bibinfo {year} {2023})},\
  \Eprint {http://arxiv.org/abs/2304.14496} {arXiv:2304.14496
  [physics.ins-det]} \BibitemShut {NoStop}%
\bibitem [{\citenamefont {Lieret}\ \emph {et~al.}(2023)\citenamefont {Lieret},
  \citenamefont {DeZoort}, \citenamefont {Chatterjee}, \citenamefont {Park},
  \citenamefont {Miao},\ and\ \citenamefont {Li}}]{Lieret:2023aqg}%
  \BibitemOpen
  \bibfield  {author} {\bibinfo {author} {\bibfnamefont {Kilian}\ \bibnamefont
  {Lieret}}, \bibinfo {author} {\bibfnamefont {Gage}\ \bibnamefont {DeZoort}},
  \bibinfo {author} {\bibfnamefont {Devdoot}\ \bibnamefont {Chatterjee}},
  \bibinfo {author} {\bibfnamefont {Jian}\ \bibnamefont {Park}}, \bibinfo
  {author} {\bibfnamefont {Siqi}\ \bibnamefont {Miao}}, \ and\ \bibinfo
  {author} {\bibfnamefont {Pan}\ \bibnamefont {Li}},\ }\bibfield  {title}
  {\enquote {\bibinfo {title} {{High Pileup Particle Tracking with Object
  Condensation}},}\ \ }(\bibinfo {year} {2023})\ \Eprint
  {http://arxiv.org/abs/2312.03823} {arXiv:2312.03823 [physics.data-an]}
  \BibitemShut {NoStop}%
\bibitem [{\citenamefont {Holmberg}\ \emph {et~al.}(2023)\citenamefont
  {Holmberg}, \citenamefont {Golubovic},\ and\ \citenamefont
  {Kirschenmann}}]{Holmberg:2023rfr}%
  \BibitemOpen
  \bibfield  {author} {\bibinfo {author} {\bibfnamefont {Daniel}\ \bibnamefont
  {Holmberg}}, \bibinfo {author} {\bibfnamefont {Dejan}\ \bibnamefont
  {Golubovic}}, \ and\ \bibinfo {author} {\bibfnamefont {Henning}\ \bibnamefont
  {Kirschenmann}},\ }\bibfield  {title} {\enquote {\bibinfo {title} {{Jet
  Energy Calibration with Deep Learning as a Kubeflow Pipeline}},}\ }\href
  {\doibase 10.1007/s41781-023-00103-y} {\bibfield  {journal} {\bibinfo
  {journal} {Comput. Softw. Big Sci.}\ }\textbf {\bibinfo {volume} {7}},\
  \bibinfo {pages} {9} (\bibinfo {year} {2023})},\ \Eprint
  {http://arxiv.org/abs/2308.12724} {arXiv:2308.12724 [hep-ex]} \BibitemShut
  {NoStop}%
\bibitem [{\citenamefont {Sirunyan}\ \emph {et~al.}(2019)\citenamefont
  {Sirunyan} \emph {et~al.}}]{Sirunyan:2019wwa}%
  \BibitemOpen
  \bibfield  {author} {\bibinfo {author} {\bibfnamefont {Albert~M}\
  \bibnamefont {Sirunyan}} \emph {et~al.} (\bibinfo {collaboration} {CMS}),\
  }\bibfield  {title} {\enquote {\bibinfo {title} {{A deep neural network for
  simultaneous estimation of b jet energy and resolution}},}\ }\href {\doibase
  10.1007/s41781-020-00041-z} {\  (\bibinfo {year} {2019}),\
  10.1007/s41781-020-00041-z},\ \Eprint {http://arxiv.org/abs/1912.06046}
  {arXiv:1912.06046 [hep-ex]} \BibitemShut {NoStop}%
\bibitem [{\citenamefont {Cheong}\ \emph {et~al.}(2020)\citenamefont {Cheong},
  \citenamefont {Cukierman}, \citenamefont {Nachman}, \citenamefont {Safdari},\
  and\ \citenamefont {Schwartzman}}]{Cheong:2019upg}%
  \BibitemOpen
  \bibfield  {author} {\bibinfo {author} {\bibfnamefont {Sanha}\ \bibnamefont
  {Cheong}}, \bibinfo {author} {\bibfnamefont {Aviv}\ \bibnamefont
  {Cukierman}}, \bibinfo {author} {\bibfnamefont {Benjamin}\ \bibnamefont
  {Nachman}}, \bibinfo {author} {\bibfnamefont {Murtaza}\ \bibnamefont
  {Safdari}}, \ and\ \bibinfo {author} {\bibfnamefont {Ariel}\ \bibnamefont
  {Schwartzman}},\ }\bibfield  {title} {\enquote {\bibinfo {title}
  {{Parametrizing the Detector Response with Neural Networks}},}\ }\href
  {\doibase 10.1088/1748-0221/15/01/P01030} {\bibfield  {journal} {\bibinfo
  {journal} {JINST}\ }\textbf {\bibinfo {volume} {15}},\ \bibinfo {pages}
  {P01030} (\bibinfo {year} {2020})},\ \Eprint
  {http://arxiv.org/abs/1910.03773} {arXiv:1910.03773 [physics.data-an]}
  \BibitemShut {NoStop}%
\bibitem [{\citenamefont {Bollweg}\ \emph {et~al.}(2020)\citenamefont
  {Bollweg}, \citenamefont {Haußmann}, \citenamefont {Kasieczka},
  \citenamefont {Luchmann}, \citenamefont {Plehn},\ and\ \citenamefont
  {Thompson}}]{Bollweg:2019skg}%
  \BibitemOpen
  \bibfield  {author} {\bibinfo {author} {\bibfnamefont {Sven}\ \bibnamefont
  {Bollweg}}, \bibinfo {author} {\bibfnamefont {Manuel}\ \bibnamefont
  {Haußmann}}, \bibinfo {author} {\bibfnamefont {Gregor}\ \bibnamefont
  {Kasieczka}}, \bibinfo {author} {\bibfnamefont {Michel}\ \bibnamefont
  {Luchmann}}, \bibinfo {author} {\bibfnamefont {Tilman}\ \bibnamefont
  {Plehn}}, \ and\ \bibinfo {author} {\bibfnamefont {Jennifer}\ \bibnamefont
  {Thompson}},\ }\bibfield  {title} {\enquote {\bibinfo {title} {{Deep-Learning
  Jets with Uncertainties and More}},}\ }\href {\doibase
  10.21468/SciPostPhys.8.1.006} {\bibfield  {journal} {\bibinfo  {journal}
  {SciPost Phys.}\ }\textbf {\bibinfo {volume} {8}},\ \bibinfo {pages} {006}
  (\bibinfo {year} {2020})},\ \Eprint {http://arxiv.org/abs/1904.10004}
  {arXiv:1904.10004 [hep-ph]} \BibitemShut {NoStop}%
\bibitem [{\citenamefont {Bellagente}\ \emph {et~al.}(2021)\citenamefont
  {Bellagente}, \citenamefont {Hau\ss{}mann}, \citenamefont {Luchmann},\ and\
  \citenamefont {Plehn}}]{Bellagente:2021yyh}%
  \BibitemOpen
  \bibfield  {author} {\bibinfo {author} {\bibfnamefont {Marco}\ \bibnamefont
  {Bellagente}}, \bibinfo {author} {\bibfnamefont {Manuel}\ \bibnamefont
  {Hau\ss{}mann}}, \bibinfo {author} {\bibfnamefont {Michel}\ \bibnamefont
  {Luchmann}}, \ and\ \bibinfo {author} {\bibfnamefont {Tilman}\ \bibnamefont
  {Plehn}},\ }\bibfield  {title} {\enquote {\bibinfo {title} {{Understanding
  Event-Generation Networks via Uncertainties}},}\ }\href@noop {} {\  (\bibinfo
  {year} {2021})},\ \Eprint {http://arxiv.org/abs/2104.04543} {arXiv:2104.04543
  [hep-ph]} \BibitemShut {NoStop}%
\bibitem [{\citenamefont {Araz}\ and\ \citenamefont
  {Spannowsky}(2021)}]{Araz:2021wqm}%
  \BibitemOpen
  \bibfield  {author} {\bibinfo {author} {\bibfnamefont {Jack~Y.}\ \bibnamefont
  {Araz}}\ and\ \bibinfo {author} {\bibfnamefont {Michael}\ \bibnamefont
  {Spannowsky}},\ }\bibfield  {title} {\enquote {\bibinfo {title} {{Combine and
  Conquer: Event Reconstruction with Bayesian Ensemble Neural Networks}},}\
  }\href@noop {} {\  (\bibinfo {year} {2021})},\ \Eprint
  {http://arxiv.org/abs/2102.01078} {arXiv:2102.01078 [hep-ph]} \BibitemShut
  {NoStop}%
\bibitem [{\citenamefont {Kronheim}\ \emph {et~al.}(2021)\citenamefont
  {Kronheim}, \citenamefont {Kuchera}, \citenamefont {Prosper},\ and\
  \citenamefont {Ramanujan}}]{Kronheim:2021hdb}%
  \BibitemOpen
  \bibfield  {author} {\bibinfo {author} {\bibfnamefont {Braden}\ \bibnamefont
  {Kronheim}}, \bibinfo {author} {\bibfnamefont {Michelle~P.}\ \bibnamefont
  {Kuchera}}, \bibinfo {author} {\bibfnamefont {Harrison~B.}\ \bibnamefont
  {Prosper}}, \ and\ \bibinfo {author} {\bibfnamefont {Raghuram}\ \bibnamefont
  {Ramanujan}},\ }\bibfield  {title} {\enquote {\bibinfo {title} {{Implicit
  Quantile Neural Networks for Jet Simulation and Correction}},}\ }\href@noop
  {} {\  (\bibinfo {year} {2021})},\ \Eprint {http://arxiv.org/abs/2111.11415}
  {arXiv:2111.11415 [physics.comp-ph]} \BibitemShut {NoStop}%
\bibitem [{cer()}]{cernopendata}%
  \BibitemOpen
  \href {http://opendata.cern.ch} {\enquote {\bibinfo {title} {Cern open data
  portal},}\ }\BibitemShut {NoStop}%
\bibitem [{\citenamefont {Aaboud}\ \emph {et~al.}(2019)\citenamefont {Aaboud}
  \emph {et~al.}}]{ATLAS:2018fwq}%
  \BibitemOpen
  \bibfield  {author} {\bibinfo {author} {\bibfnamefont {Morad}\ \bibnamefont
  {Aaboud}} \emph {et~al.} (\bibinfo {collaboration} {ATLAS}),\ }\bibfield
  {title} {\enquote {\bibinfo {title} {{Measurement of the top quark mass in
  the $t\bar{t}\rightarrow $ lepton+jets channel from $\sqrt{s}=8$ TeV ATLAS
  data and combination with previous results}},}\ }\href {\doibase
  10.1140/epjc/s10052-019-6757-9} {\bibfield  {journal} {\bibinfo  {journal}
  {Eur. Phys. J. C}\ }\textbf {\bibinfo {volume} {79}},\ \bibinfo {pages} {290}
  (\bibinfo {year} {2019})},\ \Eprint {http://arxiv.org/abs/1810.01772}
  {arXiv:1810.01772 [hep-ex]} \BibitemShut {NoStop}%
\bibitem [{\citenamefont {Tumasyan}\ \emph {et~al.}(2023)\citenamefont
  {Tumasyan} \emph {et~al.}}]{CMS:2023ebf}%
  \BibitemOpen
  \bibfield  {author} {\bibinfo {author} {\bibfnamefont {Armen}\ \bibnamefont
  {Tumasyan}} \emph {et~al.} (\bibinfo {collaboration} {CMS}),\ }\bibfield
  {title} {\enquote {\bibinfo {title} {{Measurement of the top quark mass using
  a profile likelihood approach with the lepton~+~jets final states in
  proton\textendash{}proton collisions at $\sqrt{s}=13\,\text
  {Te}\hspace{-.08em}\text {V} $}},}\ }\href {\doibase
  10.1140/epjc/s10052-023-12050-4} {\bibfield  {journal} {\bibinfo  {journal}
  {Eur. Phys. J. C}\ }\textbf {\bibinfo {volume} {83}},\ \bibinfo {pages} {963}
  (\bibinfo {year} {2023})},\ \Eprint {http://arxiv.org/abs/2302.01967}
  {arXiv:2302.01967 [hep-ex]} \BibitemShut {NoStop}%
\bibitem [{\citenamefont
  {Tibshirani}(1996)}]{51791361-8fe2-38d5-959f-ae8d048b490d}%
  \BibitemOpen
  \bibfield  {author} {\bibinfo {author} {\bibfnamefont {Robert}\ \bibnamefont
  {Tibshirani}},\ }\bibfield  {title} {\enquote {\bibinfo {title} {Regression
  shrinkage and selection via the lasso},}\ }\href
  {http://www.jstor.org/stable/2346178} {\bibfield  {journal} {\bibinfo
  {journal} {Journal of the Royal Statistical Society. Series B
  (Methodological)}\ }\textbf {\bibinfo {volume} {58}},\ \bibinfo {pages}
  {267--288} (\bibinfo {year} {1996})}\BibitemShut {NoStop}%
\bibitem [{\citenamefont {Chatrchyan}\ \emph {et~al.}(2008)\citenamefont
  {Chatrchyan} \emph {et~al.}}]{CMS:2008xjf}%
  \BibitemOpen
  \bibfield  {author} {\bibinfo {author} {\bibfnamefont {S.}~\bibnamefont
  {Chatrchyan}} \emph {et~al.} (\bibinfo {collaboration} {CMS}),\ }\bibfield
  {title} {\enquote {\bibinfo {title} {{The CMS Experiment at the CERN LHC}},}\
  }\href {\doibase 10.1088/1748-0221/3/08/S08004} {\bibfield  {journal}
  {\bibinfo  {journal} {JINST}\ }\textbf {\bibinfo {volume} {3}},\ \bibinfo
  {pages} {S08004} (\bibinfo {year} {2008})}\BibitemShut {NoStop}%
\bibitem [{\citenamefont {Sjöstrand}\ \emph {et~al.}(2006)\citenamefont
  {Sjöstrand}, \citenamefont {Mrenna},\ and\ \citenamefont
  {Skands}}]{Sj_strand_2006}%
  \BibitemOpen
  \bibfield  {author} {\bibinfo {author} {\bibfnamefont {Torbjörn}\
  \bibnamefont {Sjöstrand}}, \bibinfo {author} {\bibfnamefont {Stephen}\
  \bibnamefont {Mrenna}}, \ and\ \bibinfo {author} {\bibfnamefont {Peter}\
  \bibnamefont {Skands}},\ }\bibfield  {title} {\enquote {\bibinfo {title}
  {{PYTHIA} 6.4 physics and manual},}\ }\href {\doibase
  10.1088/1126-6708/2006/05/026} {\bibfield  {journal} {\bibinfo  {journal}
  {Journal of High Energy Physics}\ }\textbf {\bibinfo {volume} {2006}},\
  \bibinfo {pages} {026--026} (\bibinfo {year} {2006})}\BibitemShut {NoStop}%
\bibitem [{\citenamefont {Agostinelli}\ \emph {et~al.}(2003)\citenamefont
  {Agostinelli} \emph {et~al.}}]{AGOSTINELLI2003250}%
  \BibitemOpen
  \bibfield  {author} {\bibinfo {author} {\bibfnamefont {S.}~\bibnamefont
  {Agostinelli}} \emph {et~al.},\ }\bibfield  {title} {\enquote {\bibinfo
  {title} {Geant4—a simulation toolkit},}\ }\href {\doibase
  https://doi.org/10.1016/S0168-9002(03)01368-8} {\bibfield  {journal}
  {\bibinfo  {journal} {Nuclear Instruments and Methods in Physics Research
  Section A: Accelerators, Spectrometers, Detectors and Associated Equipment}\
  }\textbf {\bibinfo {volume} {506}},\ \bibinfo {pages} {250--303} (\bibinfo
  {year} {2003})}\BibitemShut {NoStop}%
\bibitem [{\citenamefont {Komiske}\ \emph
  {et~al.}(2019{\natexlab{a}})\citenamefont {Komiske}, \citenamefont
  {Mastandrea}, \citenamefont {Metodiev}, \citenamefont {Naik},\ and\
  \citenamefont {Thaler}}]{komiske_patrick_2019_3340205}%
  \BibitemOpen
  \bibfield  {author} {\bibinfo {author} {\bibfnamefont {Patrick}\ \bibnamefont
  {Komiske}}, \bibinfo {author} {\bibfnamefont {Radha}\ \bibnamefont
  {Mastandrea}}, \bibinfo {author} {\bibfnamefont {Eric}\ \bibnamefont
  {Metodiev}}, \bibinfo {author} {\bibfnamefont {Preksha}\ \bibnamefont
  {Naik}}, \ and\ \bibinfo {author} {\bibfnamefont {Jesse}\ \bibnamefont
  {Thaler}},\ }\href {\doibase 10.5281/zenodo.3340205} {\enquote {\bibinfo
  {title} {{CMS 2011A Open Data | Jet Primary Dataset | pT > 375 GeV | MOD HDF5
  Format}},}\ } (\bibinfo {year} {2019}{\natexlab{a}})\BibitemShut {NoStop}%
\bibitem [{\citenamefont {Beaudette}(2013)}]{Beaudette:2013kbl}%
  \BibitemOpen
  \bibfield  {author} {\bibinfo {author} {\bibfnamefont {Florian}\ \bibnamefont
  {Beaudette}} (\bibinfo {collaboration} {CMS}),\ }\bibfield  {title} {\enquote
  {\bibinfo {title} {{The CMS Particle Flow Algorithm}},}\ }in\ \href@noop {}
  {\emph {\bibinfo {booktitle} {{International Conference on Calorimetry for
  the High Energy Frontier}}}}\ (\bibinfo {year} {2013})\ pp.\ \bibinfo {pages}
  {295--304},\ \Eprint {http://arxiv.org/abs/1401.8155} {arXiv:1401.8155
  [hep-ex]} \BibitemShut {NoStop}%
\bibitem [{\citenamefont {Cacciari}\ and\ \citenamefont
  {Salam}(2006)}]{Cacciari:2005hq}%
  \BibitemOpen
  \bibfield  {author} {\bibinfo {author} {\bibfnamefont {Matteo}\ \bibnamefont
  {Cacciari}}\ and\ \bibinfo {author} {\bibfnamefont {Gavin~P.}\ \bibnamefont
  {Salam}},\ }\bibfield  {title} {\enquote {\bibinfo {title} {{Dispelling the
  $N^{3}$ myth for the $k_t$ jet-finder}},}\ }\href {\doibase
  10.1016/j.physletb.2006.08.037} {\bibfield  {journal} {\bibinfo  {journal}
  {Phys. Lett.}\ }\textbf {\bibinfo {volume} {B641}},\ \bibinfo {pages} {57}
  (\bibinfo {year} {2006})},\ \Eprint {http://arxiv.org/abs/hep-ph/0512210}
  {arXiv:hep-ph/0512210 [hep-ph]} \BibitemShut {NoStop}%
%%CITATION = HEP-PH/0512210;%%
\bibitem [{\citenamefont {Cacciari}\ \emph {et~al.}(2008)\citenamefont
  {Cacciari}, \citenamefont {Salam},\ and\ \citenamefont
  {Soyez}}]{Cacciari_2008}%
  \BibitemOpen
  \bibfield  {author} {\bibinfo {author} {\bibfnamefont {Matteo}\ \bibnamefont
  {Cacciari}}, \bibinfo {author} {\bibfnamefont {Gavin~P}\ \bibnamefont
  {Salam}}, \ and\ \bibinfo {author} {\bibfnamefont {Gregory}\ \bibnamefont
  {Soyez}},\ }\bibfield  {title} {\enquote {\bibinfo {title} {The anti-ktjet
  clustering algorithm},}\ }\href {\doibase 10.1088/1126-6708/2008/04/063}
  {\bibfield  {journal} {\bibinfo  {journal} {Journal of High Energy Physics}\
  }\textbf {\bibinfo {volume} {2008}},\ \bibinfo {pages} {063–063} (\bibinfo
  {year} {2008})}\BibitemShut {NoStop}%
\bibitem [{\citenamefont {Cacciari}\ \emph {et~al.}(2012)\citenamefont
  {Cacciari}, \citenamefont {Salam},\ and\ \citenamefont
  {Soyez}}]{Cacciari:2011ma}%
  \BibitemOpen
  \bibfield  {author} {\bibinfo {author} {\bibfnamefont {Matteo}\ \bibnamefont
  {Cacciari}}, \bibinfo {author} {\bibfnamefont {Gavin~P.}\ \bibnamefont
  {Salam}}, \ and\ \bibinfo {author} {\bibfnamefont {Gregory}\ \bibnamefont
  {Soyez}},\ }\bibfield  {title} {\enquote {\bibinfo {title} {{FastJet User
  Manual}},}\ }\href {\doibase 10.1140/epjc/s10052-012-1896-2} {\bibfield
  {journal} {\bibinfo  {journal} {Eur. Phys. J.}\ }\textbf {\bibinfo {volume}
  {C72}},\ \bibinfo {pages} {1896} (\bibinfo {year} {2012})},\ \Eprint
  {http://arxiv.org/abs/1111.6097} {arXiv:1111.6097 [hep-ph]} \BibitemShut
  {NoStop}%
%%CITATION = ARXIV:1111.6097;%%
\bibitem [{CMS(2010)}]{CMS:2010xta}%
  \BibitemOpen
  \bibfield  {title} {\enquote {\bibinfo {title} {{Jet Performance in pp
  Collisions at 7 TeV}},}\ }\href@noop {} {\  (\bibinfo {year}
  {2010})}\BibitemShut {NoStop}%
\bibitem [{\citenamefont {Boussarie}\ \emph {et~al.}(2023)\citenamefont
  {Boussarie} \emph {et~al.}}]{Boussarie:2023izj}%
  \BibitemOpen
  \bibfield  {author} {\bibinfo {author} {\bibfnamefont {Renaud}\ \bibnamefont
  {Boussarie}} \emph {et~al.},\ }\bibfield  {title} {\enquote {\bibinfo {title}
  {{TMD Handbook}},}\ }\href@noop {} {\  (\bibinfo {year} {2023})},\ \Eprint
  {http://arxiv.org/abs/2304.03302} {arXiv:2304.03302 [hep-ph]} \BibitemShut
  {NoStop}%
\bibitem [{\citenamefont {Gallicchio}\ and\ \citenamefont
  {Schwartz}(2010)}]{Gallicchio:2010sw}%
  \BibitemOpen
  \bibfield  {author} {\bibinfo {author} {\bibfnamefont {Jason}\ \bibnamefont
  {Gallicchio}}\ and\ \bibinfo {author} {\bibfnamefont {Matthew~D.}\
  \bibnamefont {Schwartz}},\ }\bibfield  {title} {\enquote {\bibinfo {title}
  {{Seeing in Color: Jet Superstructure}},}\ }\href {\doibase
  10.1103/PhysRevLett.105.022001} {\bibfield  {journal} {\bibinfo  {journal}
  {Phys. Rev. Lett.}\ }\textbf {\bibinfo {volume} {105}},\ \bibinfo {pages}
  {022001} (\bibinfo {year} {2010})},\ \Eprint {http://arxiv.org/abs/1001.5027}
  {arXiv:1001.5027 [hep-ph]} \BibitemShut {NoStop}%
\bibitem [{\citenamefont {Abazov}\ \emph {et~al.}(2011)\citenamefont {Abazov}
  \emph {et~al.}}]{D0:2011lzz}%
  \BibitemOpen
  \bibfield  {author} {\bibinfo {author} {\bibfnamefont {Victor~Mukhamedovich}\
  \bibnamefont {Abazov}} \emph {et~al.} (\bibinfo {collaboration} {D0}),\
  }\bibfield  {title} {\enquote {\bibinfo {title} {{Measurement of Color Flow
  in $\mathbf{t\bar{t}}$ Events from $\mathbf{p\bar{p}}$ Collisions at
  $\mathbf{\sqrt{s}=1.96}$ TeV}},}\ }\href {\doibase
  10.1103/PhysRevD.83.092002} {\bibfield  {journal} {\bibinfo  {journal} {Phys.
  Rev. D}\ }\textbf {\bibinfo {volume} {83}},\ \bibinfo {pages} {092002}
  (\bibinfo {year} {2011})},\ \Eprint {http://arxiv.org/abs/1101.0648}
  {arXiv:1101.0648 [hep-ex]} \BibitemShut {NoStop}%
\bibitem [{\citenamefont {Aad}\ \emph {et~al.}(2015)\citenamefont {Aad} \emph
  {et~al.}}]{ATLAS:2015ytt}%
  \BibitemOpen
  \bibfield  {author} {\bibinfo {author} {\bibfnamefont {Georges}\ \bibnamefont
  {Aad}} \emph {et~al.} (\bibinfo {collaboration} {ATLAS}),\ }\bibfield
  {title} {\enquote {\bibinfo {title} {{Measurement of colour flow with the jet
  pull angle in $t\bar{t}$ events using the ATLAS detector at $\sqrt{s}=8$
  TeV}},}\ }\href {\doibase 10.1016/j.physletb.2015.09.051} {\bibfield
  {journal} {\bibinfo  {journal} {Phys. Lett. B}\ }\textbf {\bibinfo {volume}
  {750}},\ \bibinfo {pages} {475--493} (\bibinfo {year} {2015})},\ \Eprint
  {http://arxiv.org/abs/1506.05629} {arXiv:1506.05629 [hep-ex]} \BibitemShut
  {NoStop}%
\bibitem [{\citenamefont {Aaboud}\ \emph {et~al.}(2018)\citenamefont {Aaboud}
  \emph {et~al.}}]{ATLAS:2018olo}%
  \BibitemOpen
  \bibfield  {author} {\bibinfo {author} {\bibfnamefont {Morad}\ \bibnamefont
  {Aaboud}} \emph {et~al.} (\bibinfo {collaboration} {ATLAS}),\ }\bibfield
  {title} {\enquote {\bibinfo {title} {{Measurement of colour flow using
  jet-pull observables in $t\bar{t}$ events with the ATLAS experiment at
  $\sqrt{s} = 13\,\hbox {TeV}$}},}\ }\href {\doibase
  10.1140/epjc/s10052-018-6290-2} {\bibfield  {journal} {\bibinfo  {journal}
  {Eur. Phys. J. C}\ }\textbf {\bibinfo {volume} {78}},\ \bibinfo {pages} {847}
  (\bibinfo {year} {2018})},\ \Eprint {http://arxiv.org/abs/1805.02935}
  {arXiv:1805.02935 [hep-ex]} \BibitemShut {NoStop}%
\bibitem [{\citenamefont {Cavallini}\ \emph {et~al.}(2022)\citenamefont
  {Cavallini}, \citenamefont {Coccaro}, \citenamefont {Khosa}, \citenamefont
  {Manco}, \citenamefont {Marzani}, \citenamefont {Parodi}, \citenamefont
  {Rebuzzi}, \citenamefont {Rescia},\ and\ \citenamefont
  {Stagnitto}}]{Cavallini:2021vot}%
  \BibitemOpen
  \bibfield  {author} {\bibinfo {author} {\bibfnamefont {Luca}\ \bibnamefont
  {Cavallini}}, \bibinfo {author} {\bibfnamefont {Andrea}\ \bibnamefont
  {Coccaro}}, \bibinfo {author} {\bibfnamefont {Charanjit~K.}\ \bibnamefont
  {Khosa}}, \bibinfo {author} {\bibfnamefont {Giulia}\ \bibnamefont {Manco}},
  \bibinfo {author} {\bibfnamefont {Simone}\ \bibnamefont {Marzani}}, \bibinfo
  {author} {\bibfnamefont {Fabrizio}\ \bibnamefont {Parodi}}, \bibinfo {author}
  {\bibfnamefont {Daniela}\ \bibnamefont {Rebuzzi}}, \bibinfo {author}
  {\bibfnamefont {Alberto}\ \bibnamefont {Rescia}}, \ and\ \bibinfo {author}
  {\bibfnamefont {Giovanni}\ \bibnamefont {Stagnitto}},\ }\bibfield  {title}
  {\enquote {\bibinfo {title} {{Tagging the Higgs boson decay to bottom quarks
  with colour-sensitive observables and the Lund jet plane}},}\ }\href
  {\doibase 10.1140/epjc/s10052-022-10447-1} {\bibfield  {journal} {\bibinfo
  {journal} {Eur. Phys. J. C}\ }\textbf {\bibinfo {volume} {82}},\ \bibinfo
  {pages} {493} (\bibinfo {year} {2022})},\ \Eprint
  {http://arxiv.org/abs/2112.09650} {arXiv:2112.09650 [hep-ph]} \BibitemShut
  {NoStop}%
\bibitem [{\citenamefont {Komiske}\ \emph
  {et~al.}(2019{\natexlab{b}})\citenamefont {Komiske}, \citenamefont
  {Mastandrea}, \citenamefont {Metodiev}, \citenamefont {Naik},\ and\
  \citenamefont {Thaler}}]{komiske_patrick_2019_3341502}%
  \BibitemOpen
  \bibfield  {author} {\bibinfo {author} {\bibfnamefont {Patrick}\ \bibnamefont
  {Komiske}}, \bibinfo {author} {\bibfnamefont {Radha}\ \bibnamefont
  {Mastandrea}}, \bibinfo {author} {\bibfnamefont {Eric}\ \bibnamefont
  {Metodiev}}, \bibinfo {author} {\bibfnamefont {Preksha}\ \bibnamefont
  {Naik}}, \ and\ \bibinfo {author} {\bibfnamefont {Jesse}\ \bibnamefont
  {Thaler}},\ }\href {\doibase 10.5281/zenodo.3341502} {\enquote {\bibinfo
  {title} {{CMS 2011A Simulation | Pythia 6 QCD 1000-1400 | pT > 375 GeV | MOD
  HDF5 Format}},}\ } (\bibinfo {year} {2019}{\natexlab{b}})\BibitemShut
  {NoStop}%
\bibitem [{\citenamefont {Komiske}\ \emph
  {et~al.}(2019{\natexlab{c}})\citenamefont {Komiske}, \citenamefont
  {Mastandrea}, \citenamefont {Metodiev}, \citenamefont {Naik},\ and\
  \citenamefont {Thaler}}]{komiske_patrick_2019_3341770}%
  \BibitemOpen
  \bibfield  {author} {\bibinfo {author} {\bibfnamefont {Patrick}\ \bibnamefont
  {Komiske}}, \bibinfo {author} {\bibfnamefont {Radha}\ \bibnamefont
  {Mastandrea}}, \bibinfo {author} {\bibfnamefont {Eric}\ \bibnamefont
  {Metodiev}}, \bibinfo {author} {\bibfnamefont {Preksha}\ \bibnamefont
  {Naik}}, \ and\ \bibinfo {author} {\bibfnamefont {Jesse}\ \bibnamefont
  {Thaler}},\ }\href {\doibase 10.5281/zenodo.3341770} {\enquote {\bibinfo
  {title} {{CMS 2011A Simulation | Pythia 6 QCD 1400-1800 | pT > 375 GeV | MOD
  HDF5 Format}},}\ } (\bibinfo {year} {2019}{\natexlab{c}})\BibitemShut
  {NoStop}%
\bibitem [{\citenamefont {Komiske}\ \emph
  {et~al.}(2019{\natexlab{d}})\citenamefont {Komiske}, \citenamefont
  {Mastandrea}, \citenamefont {Metodiev}, \citenamefont {Naik},\ and\
  \citenamefont {Thaler}}]{komiske_patrick_2019_3341772}%
  \BibitemOpen
  \bibfield  {author} {\bibinfo {author} {\bibfnamefont {Patrick}\ \bibnamefont
  {Komiske}}, \bibinfo {author} {\bibfnamefont {Radha}\ \bibnamefont
  {Mastandrea}}, \bibinfo {author} {\bibfnamefont {Eric}\ \bibnamefont
  {Metodiev}}, \bibinfo {author} {\bibfnamefont {Preksha}\ \bibnamefont
  {Naik}}, \ and\ \bibinfo {author} {\bibfnamefont {Jesse}\ \bibnamefont
  {Thaler}},\ }\href {\doibase 10.5281/zenodo.3341772} {\enquote {\bibinfo
  {title} {{CMS 2011A Simulation | Pythia 6 QCD1800-inf | pT > 375 GeV | MOD
  HDF5 Format}},}\ } (\bibinfo {year} {2019}{\natexlab{d}})\BibitemShut
  {NoStop}%
\bibitem [{\citenamefont {Nachman}\ and\ \citenamefont
  {Thaler}(2021)}]{nachman_benjamin_2021_5108967}%
  \BibitemOpen
  \bibfield  {author} {\bibinfo {author} {\bibfnamefont {Benjamin}\
  \bibnamefont {Nachman}}\ and\ \bibinfo {author} {\bibfnamefont {Jesse}\
  \bibnamefont {Thaler}},\ }\href {\doibase 10.5281/zenodo.5108967} {\enquote
  {\bibinfo {title} {Delphes dijet dataset},}\ } (\bibinfo {year}
  {2021})\BibitemShut {NoStop}%
\end{thebibliography}%

\end{document}